\documentclass[10pt,twocolumn,letterpaper]{IEEEtran}
\usepackage{amsmath,epsfig}
\usepackage{algorithm}
\usepackage{bm}
\usepackage{tikz}
\usepackage[usenames,dvipsnames]{pstricks}
\usepackage{epsfig}
\usepackage{pst-grad} 
\usepackage{pst-plot} 
\usepackage{subfigure}
\usepackage[capitalize]{cleveref}

\usepackage{subfigure}
\usepackage[T1]{fontenc}
\usepackage{amsmath}
\usepackage{graphics}
\include{psfig}
\usepackage{epsfig}
\usepackage{array}
\usepackage{psfrag}
\usepackage{amssymb}
\usepackage{color}
\usepackage{pstricks,pst-node,pst-text,pst-3d,pst-plot}
\usepackage{cite}

\usepackage{ulem}
\definecolor{dgreen}{rgb}{0, 0.8, 0.1}

\def\bz{{\bf z}}

\def\by{{\bf y}}

\def\bu{{\bf u}}

\def\bb{{\bf b}}

\def\bv{{\bf v}}

\def\bw{{\bf w}}
\def\bR{{\bf R}}
\def\br{{\bf r}}
\def\bI{{\bf I}}
\def\bE{{\bf E}}

\def\bv{{\bf v}}
\def\bA{{\bf A}}
\def\ba{{\bf a}}

\def\bs{{\bf s}}

\begin{document}


\title{Sparse Based Super Resolution Multilayer Ultrasonic Array Imaging}

\author{Shahrokh Hamidi$^*$\thanks{Shahrokh Hamidi is with the Faculty of Electrical and Computer
Engineering, University of Waterloo, 200 University Ave W, Waterloo, ON, Canada, N2L 3G1.
e-mail: \texttt{Shahrokh.Hamidi@uwaterloo.ca}.}}

\maketitle

\thispagestyle{empty}
\pagestyle{plain} 

\begin{tikzpicture}[remember picture, overlay]
      \node[font=\small] at ([yshift=-1cm]current page.north)  {\copyright IEEE};
\end{tikzpicture}

\begin{abstract}

In this paper, we model the signal propagation effect in ultrasonic imaging using Huygens principle and use this model to develop sparse signal representation based imaging techniques for a two-layer object immersed in water.
Relying on the fact that the image of interest is sparse, we cast such an array based imaging problem as a sparse signal recovery problem and develop two types of imaging methods, one method uses only one transducer to illuminate the region of interest {and for this case the system is modeled as a single input multiple output (SIMO) system. The second method relies on all transducers to transmit ultrasonic waves into the material under test and in this case the system is modeled as a multiple input multiple output (MIMO) system}. We further extend our work to a scenario where the propagation velocity of the wave in the object under test is not known precisely. {We discuss different techniques such as greedy based algorithms as well as $\ell_1$-norm minimization based approach to solve the proposed sparse signal representation based method. We give an assessment of the computational complexity of the $\ell_1$-norm minimization based approach for the SIMO and the MIMO cases.  We further point out the superiority of the $\ell_1$-norm minimization based approach over the greedy based algorithms. Then we give a comprehensive analysis of error for both the greedy based approaches as well as the $\ell_1$-norm minimization based technique for both the SIMO and the MIMO cases. The analysis utilizes tools from two powerful branches of modern analysis,  \emph{local analysis in Banach spaces} and \emph{concentration of  measure}}. We finally apply our methods to experimental data gathered from a solid test sample immersed in water and show that sparse signal recovery  based techniques outperform the conventional methods available in the literature.

\end{abstract}

\begin{IEEEkeywords}
 Ultrasonic imaging, sparse signal representation, array processing, super resolution, Huygens principle.
\end{IEEEkeywords}


\section{Introduction}
In non-destructive testing (NDT), the goal is to inspect the internal structure of materials without causing damage to them. A widely used NDT\ technique is ultrasonic array imaging, where an array of transducers is used to obtain an image of the material under test. There are two different types of ultrasonic imaging approaches, namely contact test and immersion test. In a contact test, the transducers array is in contact with test sample. In situations where the surface of the test sample is not smooth, contact test may  not be possible. In such situations, one can conduct an immersion test, where the test sample and the transducer array are immersed in a liquid such as water.
In immersion test, the gap between the transducer array and the test sample is filled with water.

 In ultrasonic array imaging, the knowledge of the array spatial signature for every point inside the region of interest  (i.e., the vector of array response to a hypothetical source located at that point) is essential to the imaging process.  This spatial signature depends on the geometry of the test setup and on properties of the environment through which the wave travels. In a homogenous medium, where the wave velocity is constant, modeling the array signature is rather straightforward. {In} non-homogenous media, where the wave velocity changes along the wave travel path, modeling the array spatial signature is not straightforward. One example of such non-homogenous media is immersion test. Indeed, the main challenge in immersion test  is the different velocities that wave experiences while passing through different layers (here water and the specimen). Due to this difference in the wave velocities in the two layers, the wave does not follow a straight line, when entering from one medium to another one. In fact, when crossing the interface between two layers, the wave is subject to refraction, which hinders the task of modeling the array spatial signature.

 One approach to account for the effect of different wave velocities is to use the so-called root mean square (RMS) velocity method which was first introduced and utilized in seismology \cite{RMS_sciesmology}. This method was applied to multi-layer ultrasonic imaging in \cite{ML_DAS}. The idea of the RMS velocity technique relies on a ray theory based approximation of the length of the path traveled by the wave, in the presence of refraction,   when it leaves from a hypothetical reflector and arrives at a transducer.
This approximation however does not take into account that wave refraction occurs  at infinitely many points on the interface between the two layers and not at one particular point on this interface. To overcome this issue, we herein use the Huygens principle to model the array spatial signature \cite{NasimPaper2}. Then, this model for the array spatial signature can be used in any imaging technique such as delay-and-sum (DAS) beamforming method, which is commonly used in ultrasonic array imaging. However, the basic shortcoming of the DAS beamformer stems from the fact that this method is independent of the statistical properties of the received data and yields poor resolution and high sidelobe levels \cite{Thesis}. It also suffers from the Rayleigh resolution limit \cite{Rayleigh} which is independent of signal-to-noise ratio (SNR). To avoid these shortcomings of the DAS beamforming approach, the  MUSIC method \cite{Music,Music_stoica,Stoica_book} and the Capon technique \cite{capon_o,Stoica_book} can be used. These techniques are highly regarded as high resolution algorithms. They outperform the DAS beamformer as they offer higher resolution and have lower sidelobe levels, while, unlike the DAS beamformer, they do not suffer from Rayleigh resolution limit. The MUSIC and Capon techniques, however, have their own disadvantages. They require to  have a large number of array snapshots  in order to obtain  a sample estimate of the covariance matrix of the array data. Furthermore, the MUSIC and Capon techniques need high value of SNR in order to yield an image with high quality. The MUSIC method is a subspace based approach and requires the knowledge of the dimension of the signal subspace (i.e., the number of the targets in the region of interest (ROI)),  which is unknown in imaging applications. Moreover, in the case of highly correlated targets both MUSIC and Capon techniques fail.

In this paper, we develop  sparse signal representation based imaging techniques which relay on the aforementioned model of the array spatial signature. Sparse signal representation based techniques have found their applications in different fields, such as synthetic aperture radar imaging \cite{SAR_application}, image reconstruction and restoration \cite{image_application}, sparse antenna array design \cite{spectrum_application},  and array processing application \cite{application_array_pro}, to mention a few. Sparse signal representation based techniques do  not suffer from Rayleigh resolution limit, their sensitivity to SNR and correlated targets are lower than that of {the MUSIC technique and the Capon method}, they can also be applied to nonlinear arrays, even with non-Gaussian measurement noises \cite{doa_sparse, Thesis, Malioutov_journal,MMV_Hyder,MMV_Cotter,MMV_Eldar}.

In this study, we use the  aforementioned model for the array spatial signature (which is based on Huygens principle)\ to   present a linear model for the array received signals. Assuming that the wave velocity in the ROI is perfectly known, we cast our imaging problem as a sparse signal recovery problem. {To do so, we first consider a case that one transducer transmits and all the receivers will receive the reflections coming back from the ROI. Therefore, we deal with a single input multiple output (SIMO) system and show how sparse signal recovery techniques can be used to image  the ROI in this case. We then address the case of multiple input multiple output (MIMO) system. Since sparse signal representation in MIMO case utilizes all the measurements, compared to the SIMO case, a better performance is expected in the sense that the sidelobe levels will be lower and   error in target location estimates will be smaller.}

{We discuss different techniques to solve the proposed sparse signal representation based method. Among the well known family of these algorithms is the greedy based algorithm which covers wide range of techniques by simple modification. We then present the $\ell_1$-norm minimization based approach and discuss the softwares that can solve it and provide an assessment of the computational complexity of the $\ell_1$-norm minimization based approach for the SIMO and  the MIMO cases.  We further point out the superiority of the $\ell_1$-norm minimization based approach over the greedy based algorithms. Then we give a comprehensive analysis of error for both the greedy based approaches as well as the $\ell_1$-norm minimization based technique for both the SIMO and the MIMO cases. The analysis utilizes tools from two powerful branches of modern analysis,  \emph{local analysis in Banach spaces} \cite{Milman,Gilles} and \emph{concentration of  measure} \cite{Measure_concentration, Ledoux}.}

We further extend our investigation and study a situation where the wave velocity in the ROI is not known. For such a situation, we reformulate our sparse signal representation based technique for the MIMO case to perform imaging without knowing the propagation velocity of the wave in the specimen, thereby introducing the so-called MIMO case for unknown velocity (MIMO-UV) algorithm. This algorithm exploits block sparsity in the image by considering a range of possible values for  the wave velocity.
We show the superiority of our sparse  signal representation based methods compared to the DAS beamformer, the MUSIC method, and the Capon technique using the experimental data gathered from a solid object immersed in water.

    In the field of ultrasonic imaging, the authors of \cite{Eldar-Ultrasound} exploit sparsity for  imaging human tissues. In such a biomedical application, there are no significant variations in the wave speed when it travels through the tissue of interest. As such, the authors of \cite{Eldar-Ultrasound} use an average figure for the wave velocity,
thereby treating the ROI, essentially,\ as a homogenous medium.
In our application of interest, the wave velocity changes abruptly when it enters from one medium to another one, and thus, the technique of  \cite{Eldar-Ultrasound} is not applicable. Another related work,  reported in  \cite{sar}, relies on the synthetic aperture array (SAA) technique. In the SAA method, only one transducer is used and data gathering and data processing is not applicable to the case where a transducer array is used. In \cite{Shahrokh}, we developed sparse signal recovery based techniques  for ultrasonic imaging of a single-layer material in the presence of mode conversion. The technique of \cite{Shahrokh} cannot be used for imaging a two-layer material.

To summarize,\textit{ our contribution in this paper is to apply sparse signal representation based techniques along with Huygens principle for imaging a  two-layer material and show their superiority, compared to the existing imaging techniques,  using experimental data. {We discuss the greedy based  algorithms as well as the $\ell_1$-norm minimization based technique as candidates to solve the proposed sparse signal representation based method. We further discuss the superiority of the $\ell_1$-norm minimization and the softwares that can solve the $\ell_1$-norm minimization problem efficiently. We also provide the computational complexity of each technique that is utilized by those softwares to solve the $\ell_1$-norm minimization problem. Moreover, we give a comprehensive analysis of error for both the greedy based approaches as well as the $\ell_1$-norm minimization based technique for both the SIMO and the MIMO cases. }}

The organization of the paper is as follows. We develop our data model in Section \ref{data_model}, and review the existing imaging methods, such as  the DAS beamformer, the MUSIC method and the Capon technique. In Section \ref{sparse_signal-representation}, we cast our problem as a sparse signal representation based problem. Then, we develop our sparse signal representation based imaging methods for  the SIMO and the MIMO cases followed by the introduction of our  imaging technique for the MIMO-UV case. {Error analysis for both the  SIMO and the MIMO cases are also given in this section}. In Section \ref{sec:exp}, we present our experimental results. Conclusions are drawn in Section \ref{conclusion}.

\section{system model}\label{data_model}
\begin{figure}
\centerline{
\includegraphics[height=5cm,width=8cm]{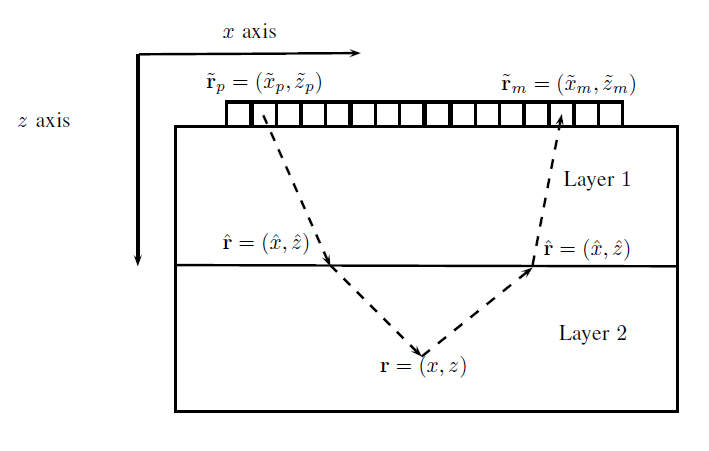}
\hspace{0.1cm}
}
\vspace*{0.1cm}
\caption{Array geometry.
\label{fig:geometry}}
\end{figure}

A linear array of $M$ transducers is used to image a two-layer material. Fig.~\ref{fig:geometry} shows the geometry of the test setup.
The wave travels with two different velocities in the two layers. In each layer,  we model the wave propagation as cylindrical wave\footnote{We herein assume that the transducers are  much longer than the depth of the test sample and thus develop our data model in a two-dimensional coordinate system. The idea of two-dimensional imaging using a one-dimensional linear array is a common practice in ultrasonic NDT,  where all quantities are assumed to be invariant in the third dimension. Our data model can be extended for three-dimensional volumetric imaging, when a two-dimensional array is employed.}. We use the Huygens principle to describe the wave in the second layer as the integral of wave field over the interface between the two layers, thereby accounting for refraction at all points on this interface. In our model, the effect of the propagation in the forward path (i.e., from the firing transducer to a scatterer) is modeled such that  this effect is combined with {scatterer's reflectivity} coefficients. Doing so, each scatterer appears to be an active source which emits the superposition {of} all waves refracted at the interface which arrived at the location of that scatterer. In what follows, we describe our date model in details.

When the $p^{\rm th}$ transducer transmits, the vector of the signals received by the array, due to back scattering, is given as{
\begin{align}
\label{data}
\by_p (\omega) = \bA(\omega;v ){\boldsymbol{\rho}_p}(\omega;v) + \bw_p(\omega)
\end{align}}
where $\by_p (\omega) \in \mathbb{C}^{M\times 1}$ contains the data received by the array when the $p^{\rm th}$ transducer transmits and is expressed as
\begin{align}
\label{received_s}
\by_p (\omega) \triangleq {[y_{1p}(\omega )\; \ y_{2p}(\omega)\; \cdots \; y_{Mp}(\omega)]^T}.
\end{align}
Here, $y_{mp}(\omega)$ is the signal received by the $m^{\rm th}$ receiver in the frequency domain, i.e., at frequency $\omega$. The signal $\bw_p(\omega) \in \mathbb{C}^{M\times 1}$ describes the vector of the receiver noises and is defined as
\begin{align}
\label{noise}
\bw_p(\omega) \triangleq [w_{1p}(\omega)\;w_{2p}(\omega)\;\cdots \; w_{Mp}(\omega)]^T
\end{align}
where $w_{mp}(\omega)$ is the noise at the output of the $m^{\rm th}$ receiver at frequency $\omega$, when the $p^{\rm th}$ transmitter transmits. Matrix $\bA(\omega;v) \in \mathbb{C}^{M\times L}$, with $L$ being the number of the point reflectors inside the second layer, is defined as
\begin{align}
\label{array_manifold}
\bA(\omega ,v)\triangleq [\ba(\omega; \br_1 ,v)\;\ba(\omega; \br_2 ,v)\;\cdots\; \ba(\omega; \br_L ,v)]\, .
\end{align}
In \eqref{array_manifold}, $v$ is the speed of the wave in the second layer and $\ba(\omega; \br ,v)\in \mathbb{C}^{M\times 1}$ is the array steering vector, corresponding to $L$ hypothetical reflectors located at $\{\br_l\}_{l=1}^{L} $, which is defined as{
\begin{align}
\label{steering vector}
\ba(\omega; \br ,v) \triangleq [ a_1(\omega; \br , v)\;a_2(\omega; \br ,v)\;\cdots\;a_M(\omega; \br ,v)]^T
\end{align}}
where $a_m(\omega; \br ,v)$, for {$m \in \{1,\cdots,M\}$}, is described as
\begin{align}
\label{steering_vector_element}
a_m(\omega; \br ,v) \triangleq \int \limits_{-\infty}^{+\infty}g_2(\omega;\hat{\br}, \br ,v)f_{21}(\omega;\tilde{\br}_m, \hat{\br})  g_1(\omega; \tilde{\br}_m, \hat{\br})\,d\hat{x}.
\end{align}
Here, $g_1(\omega; \tilde{\br}_m, \hat{\br})$ is the frequency response of the linear time invariant (LTI) system which models the signal propagation in the first layer from/to a hypothetical point located at $\hat{\br}$ that resides on the interface between the two layers to/from the $m^{\rm th}$ receiver located at $ \tilde{\br}_m$ and is given as {
\begin{align}
\label{g1}
g_1(\omega; \tilde{\br}_m,\hat{\br}) = \frac{1}{\|\tilde{\br}_m - \hat{\br}\|^{0.5}}e^{-j\omega \frac{\|\tilde{\br}_m - \hat{\br}\|}{c}}
\end{align}}
where $c$ is the speed of the wave in the first layer, {which we assume is known}. Also in (\ref{steering_vector_element}), $g_2(\omega;\mathbf{\hat{r}},\br) $ is the frequency response of the LTI system which models the signal propagation in the second layer of the test sample from/to the point scatterer located at $\br$ inside the second layer to/from the point $\mathbf{\hat{r}}$ on the interface and is expressed as
\begin{align}
\label{g2}
g_2(\omega;\hat{\br},\br ,v)= \frac{1}{\|\hat{\br} - \br\|^{0.5}}e^{-j\omega \frac{\|\hat{\br} - \br\|}{v}}.
\end{align}
Moreover in (\ref{steering_vector_element}), $f_{21}(\omega;\tilde{\br}_m,\hat{\br})$ is the transmission coefficient from Layer 2 into Layer 1, from the point located at $\hat{\br}$ on the interface toward the $m^{\rm th}$ receiver located at $\tilde{\br}_m$. According to Huygens-Fresnel principle \cite{Huygens}, we can describe $f_{21}(\omega;\tilde{\br}_m,\hat{\br})$ as
\begin{align}
\label{f21}
f_{21}(\omega;\tilde{\br}_m,\hat{\br}) = \frac{j\omega |{\hat{z}}- {\tilde{z}}_m|}{4\pi c \|\tilde{\br}_m - \hat{\br}\|}
\end{align}
{where  $\hat{z}$ is the vertical distance from the origin of the coordinate system, that we set it at the location of the first element of the array from left, to the interface. }\label{z_definition}
The vector $\boldsymbol{\rho}_p(\omega) \in \mathbb{C}^{L\times 1}$ is given as {
\begin{align}
\label{refl_coeff_vector}
\boldsymbol{\rho}_p(\omega;v) \triangleq [\rho_p(\omega; \br_1 ,v)\ \;  {\rho}_p(\omega; \br_2 ,v) \;  \cdots \; \rho_p(\omega; \br_L ,v)]^T.
\end{align}}
In (\ref{steering_vector_element}), we have described the effect of signal propagation between the point reflectors inside the second layer and the transducers. This is the signal propagation effect in the reception (or backward) path. Hence,  (\ref{steering_vector_element}) does not represent the total effect of the signal propagation. What is missing is the effect of the signal propagation between the transducer and point reflectors inside the second layer, which describes the signal propagation for the transmission (or forward) path. In fact in (\ref{refl_coeff_vector}) each {$\rho_p(\omega; \br ,v)$ for $p \in \{1,\;2,\; \cdots \;, L\}$} is expressed as  {
\begin{align}
\label{reflectivity_coeff}
& {\rho}_p(\omega; \br ,v) =  \nonumber \\ & \rho(\br) \int \limits_{-\infty}^{+\infty}g_1(\omega; \tilde{\br}_p, \hat{\br}) f_{12}(\omega;\hat{\br},\br ,v)g_2(\omega;\hat{\br},\br ,v)\,d\hat{x}
\end{align}}
where {$\rho(\br) \in \mathbb{R}^{+}$} is {a real positive number which stands for the } reflectivity coefficient of the scatterer located at $\br$ and $f_{12}(\omega;\hat{\br},\br ,v)$ is the transmission coefficient from Layer 1 to Layer 2 from point $\hat{\br}$ on the interface toward the point located at $\br$ inside the second layer. Note that in our forthcoming imaging algorithms, we are interested in finding {$\rho(\br)$} as the image of the ROI. According to the Huygens-Fresnel principle \cite{Huygens}, $f_{12}(\omega;\hat{\br}, \br ,v)$ is given by {
\begin{align}
\label{f12}
f_{12}(\omega;\hat{\br}, \br ,v)= \frac{j\omega |z-\hat{z}|}{4\pi v \|\hat{\br} - \br\|}.
\end{align}}
Thus, the signal propagation effect in the transmission path is described by the integral on the right hand sight of (\ref{reflectivity_coeff}).

Before describing our problem as a sparse signal representation based problem, we review the DAS beamformer, the MUSIC method and the Capon technique, in the subsequent subsections.

{\subsection{The DAS Beamformer}}
The image provided by the DAS beamformer, i.e., the estimate of the reflectivity coefficients for a potential reflector located at $\br$, is denoted as $\mathcal{I}_{\rm{DAS}}(\omega;\br,v)$ and is given as{
\begin{align}
\label{DAS}
\mathcal{I}_{\rm{DAS}}(\omega;\br,v) =  {\left|  \sum_{p=1}^M  \sum_{\omega \in \mathbf{\Omega}}
\ba^H(\omega;\br ,v) \by_p(\omega)\right|}^2
\end{align}}
where $\by_p(\omega)$ is given in (\ref{data}), $(\cdot)^H$ stands for conjugate transpose, $\mathbf{\Omega} $ is the set of all frequencies in the bandwidth of the probing signal, and $\ba(\omega;\br,v)$ is given as in (\ref{steering vector}). The DAS beamformer can be easily implemented as its computational complexity is relatively low. However, it suffers from high sidelobe level and low resolution \cite{Thesis}. Indeed, its resolution is bound by the Rayleigh resolution limit which is independent of SNR \cite{Rayleigh}.
{\subsection{MUSIC Based Imaging}}
To enhance the resolution and to decrease the sidelobe level, the MUSIC method \cite{Music,Music_stoica,Stoica_book} can be used for imaging. The MUSIC method is a subspace based technique which exploits the second order statistics of the received data to image the material under test. The basic building block for the MUSIC technique is the sample covariance matrix which is described as{
\begin{align}
\label{covariance_matrix}
\hat{\bR} (\omega ) = \frac{1}{M}\sum\limits_{p = 1}^M  \by_p(\omega)\by_p^H(\omega).
\end{align}}
The image provided by the MUSIC method is given as \cite{Music,Music_stoica,Stoica_book}{
\begin{align}
\label{music_all_freq}
\mathcal{I}_{\rm{MUSIC}}(\omega;\br,v)& = \sum\limits_{\omega  \in \mathbf{\Omega}} {\frac{{{\ba^H(\omega; \br,v)}\ba(\omega; \br,v)}}{{{\ba^H(\omega; \br,v)}{\bE_n}(\omega ){\bE^H_n}{{(\omega )}}\ba(\omega; \br,v)}}}
\end{align}}
where the columns of ${\bE_n}(\omega) \in \mathbb{C}^{M\times (M-L)}$ are the eigenvectors of the matrix
$\mathbf{\hat{R}} (\omega )$ corresponding to the smallest $ M-L$ eigenvalues with $L$ being the effective dimension of the signal subspace.  In order to incorporate the information from all the frequency bins, in \eqref{music_all_freq},  {we} use a summation over the bandwidth of the probing signal to add the MUSIC\ image obtained from different frequency bins. At the end, the $L$ highest peaks of these function give us the location of the $L$ reflectors.
{\subsection{Capon Based Imaging}}
Capon filter bank approach is another high resolution technique that utilizes the sample covariance matrix to generate a high resolution image for a point reflector located at {$\br$} \cite{capon_o,Stoica_book}. The Capon image at frequency $\omega$ is given as
\begin{align}
\label{Capon_all_freq}
\mathcal{I}_{\rm{Capon}}(\omega;\br,v) = \sum\limits_{\omega  \in \mathbf{\Omega}} {\frac{{{\ba^H(\omega; \br,v)}\ba(\omega; \br,v)}}{{{\ba^H(\omega; \br,v)}{\hat{\bR}_{\rm DL}^{-1}(\omega)}(\omega)\ba(\omega; \br,v)}}}\,
\end{align}
where
\begin{align}\label{eq:DL_R}
 \hat{\bR}_{\rm DL} (\omega) \triangleq   \hat  \bR (\omega)   + \kappa \bI,
\end{align}
is the so-called diagonally loaded sample covariance matrix and $\kappa$ is the so-called diagonal loading factor which is a tunable parameter to calibrate the resulting Capon image. The standard choice for $\kappa$ is $10$ to $12$ dB above the transducer noise level \cite{diagonal_loading_1,diagonal_loading_2,diagonal_loading_3}. The  location of the $L$ highest peaks of the image in (\ref{Capon_all_freq}) are introduced  as  the locations of the $L$ reflectors.
\section{Sparse Signal Representation Based Techniques}\label{sparse_signal-representation}
\subsection{Single Input Multiple Output Case}
\begin{figure}
\centerline{
\includegraphics[height=5cm,width=8cm]{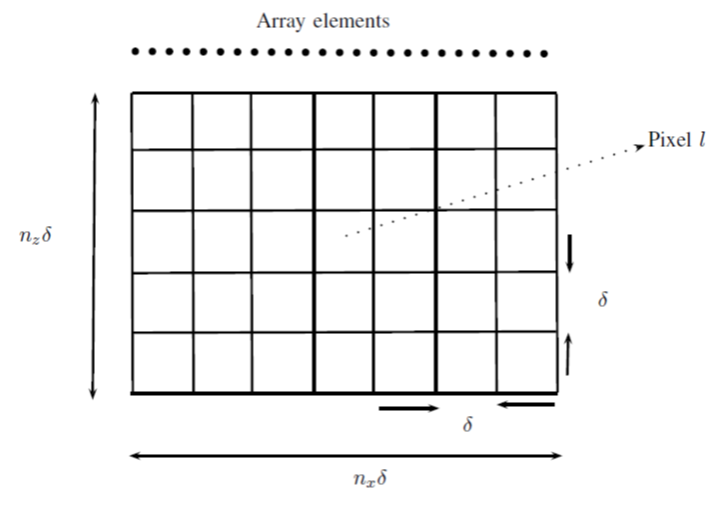}
\hspace{0.1cm}
}
\vspace*{0.1cm}
\caption{Two-dimensional grid covering the ROI.
\label{fig:Grid}}
\end{figure}

Despite all the aforementioned advantages for {the MUSIC technique and the Capon method}, they have certain drawbacks. As a subspace based approach, the MUSIC technique utilizes the signal subspace where {its} dimension is the number of the true reflectors. In practice, however, finding the number of true reflectors is not an easy task. Both {the MUSIC technique and the Capon method} are very sensitive {to} the correlation between the reflectors. Regarding the Capon method, although the resolution can be improved by increasing SNR, the sidelobe level is restricted to $\sigma ^2/M$ where $\sigma ^2$ is the power of the noise. One way to overcome these shortcomings is to use the sparsity property of the underlying image. Knowing a-priori that the desired image is sparse, one can exploit this sparsity to the advantage of the imaging process. Due to their lower sidelobe levels and higher resolution these approaches are superior compared to the DAS beamformer, the MUSIC technique, and the Capon method \cite{Thesis}. Its sensitivity to SNR and correlation between reflectors is much less than those of {the  MUSIC technique and the Capon method}.  To cast the problem as a sparse signal recovery problem, at each frequency we define $\mathbf{\Phi}(\omega;v) \in \mathbb{C}^{M\times N} $ as the dictionary matrix where $N$ represents the number of the potential reflectors {and $N>M$}. To guarantee that our problem is sparse, $N$ must be much greater than the number of {reflectors} $(N \gg L)$. To accomplish this goal, we divide the ROI into $n_x \times n_z = N$ pixels. Each pixel represents a potential reflector. Fig.~\ref{fig:Grid} shows this grid, with $n_x$ pixels in the horizontal direction and $n_z$ pixels in the vertical direction. {Note that we impose the sparsity to the imaging problem by choosing $N\gg L$. In fact when $N \gg L$, the problem is called a sparse problem. We should also point out that the only role of the Huygens principle is to find the array spatial signature.}\label{Sparsity_Huygens}

The matrix $\mathbf{\Phi}(\omega; v)\in \mathbb{C}^{M\times N}$ for {$N > M$} is defined as
\begin{align}
\label{phi}
{\mathbf{\Phi}}(\omega;v) \triangleq [\ba(\omega; \check{\br}_1 ,v)\; \ba(\omega; \check{\br}_2 ,v)\; \cdots  \;\ba(\omega; \check{\br}_N,v)].
\end{align}
In (\ref{phi}), $\check{\br}_l$ is the location of the $l^{\rm th}$ potential scatterer, $\ba(\omega; \check{\br}_l,v) \in \mathbb{C}^{M\times 1}$ is the array steering vector for the $l^{\rm th}$ potential scatterer at $\check{\br}_l$ (corresponding to the $(i,j)^{\rm th}$ pixel in the ROI where $i=\lfloor\frac{l-1}{n_x}\rfloor+1$ and $j=l-(i-1)n_x$) and it is given in (\ref{steering vector}). {It is worth mentioning that the matrix $\mathbf{\Phi}(\omega; v)$ given in (\ref{phi}) only incorporates the reception phase}\label{transmission_phase_only}. Using the dictionary defined in (\ref{phi}), our model presented in (\ref{data}) can be rewritten as{
\begin{align}
\label{model_sparse_smv}
\by_p(\omega) = {\mathbf{\Phi}}(\omega;v)\bs + \bw_p(\omega)
\end{align}}
where the $l^{\rm th}$ element of $\bs \in \mathbb{C}^{N\times 1}$ is nonzero if there is a scatterer located at the $(i,j)^{\rm th}$ pixel, where $i= \lfloor\frac{l-1}{n_x}\rfloor+1$ and $ j=l-(i-1)n_x $, and it is zero otherwise. Each nonzero element of $\bs$ is one of the entries of (\ref{refl_coeff_vector}). Our goal is to estimate $s(\check{\br}_l)$ which is the reflectivity coefficient of a point target located at $\check{\br}_l$ corresponding to the $(i,j)^{\rm th}$ pixel of the ROI.
{There are different techniques to recover the sparse signal $\bs$ for a given model in (\ref{model_sparse_smv}). Among them we can mention the greedy based techniques and the $\ell_1$-norm minimization based method. In the following we discuss these techniques.
\paragraph{Orthogonal Greedy Algorithm}
The orthogonal greedy algorithm (OGA) is a heuristic approach to find the sparsest vector $\bs$ in (\ref{model_sparse_smv}). In the OGA one solves the following optimization problem \cite{Elad,OGA}
\begin{align}
\label{OGA}
i_k = \arg\max_{1 \leq i \leq N}|<r^{(k-1)},{\mathbf{\Phi}}(\omega;v)>|
\end{align}
where $<\cdot,\cdot>$ stands for the inner product in Euclidean space and
\begin{align}
\label{residue}
r^{(k)} = \by_p(\omega)-\hat{\by}_p^{(k)}(\omega).
\end{align}
In (\ref{residue}) the term $\hat{\by}_p^{(k)}(\omega)$ is given as
\begin{align}
\hat{\by}_p^{(k)}(\omega) = \sum\limits_{i=1}^{k} \chi_{il}^k {\Phi}_{il}(\omega;v),
\end{align}
in which the coefficients $ \chi_{il}^k $ are fitted by least squares to minimize $\|\by_p(\omega)-\hat{\by}_p^{(k)}(\omega)\|^2$.
The algorithm stops when the error term $r$ falls below a predetermined threshold.
Known as forward stepwise regression, the OGA has been widely used in the setting of statistical modeling since 1960's \cite{GOA}. In signal processing, the OGA is known as matching pursuit (MP) \cite{MP}. The algorithm (\ref{OGA}) is in fact called orthogonal matching pursuit (OMP) \cite{OMP}.}\label{OGA_algorithm}

\paragraph{$\ell_1$-norm minimization based technique}
A well known technique to find the sparsest signal $\bs$ in (\ref{model_sparse_smv}) is based on the $\ell_1$-norm minimization.
Based on (\ref{model_sparse_smv}), the following $\ell_1$-norm minimization problem can be used to obtain $\bs$: {
\begin{align}
 \label{l1_SMV}
 \rm{\min_{\bs}}. \ & \| \bs\|_1 \ \nonumber \\
 \rm{subject \ to} \ & \|\by_p(\omega) - {\mathbf{\Phi}}(\omega;v ) \bs\|_2\le \beta_s.
 \end{align}}
 This optimization problem is known as basis pursuit de-noising \cite{basis_pusuit} or least absolute shrinkage and selection operator {(LASSO)} \cite{lasso} and can be solved using any software packages dedicated to solving conventional optimization problem {such as CVX software package \cite{cvx}}. {It has been shown in \cite{basis_pusuit} that (\ref{l1_SMV}) outperforms the OGA, given in (\ref{OGA}). The CVX software package casts the optimization problem in (\ref{l1_SMV}) as a second order cone programming (SOCP) and solves it using interior point method \cite{Boyd_book}.
The parameter $\beta_s$ is the regularization parameter and is chosen such that the probability of the norm of the noise vector being larger than $\beta_s$ is small \cite{Malioutov_journal,Thesis}.}\label{beta_s}
After obtaining $\bs$, we reshape it into an $n_z \times n_x$ matrix which can be used to obtain the final image. According to (\ref{refl_coeff_vector}) and (\ref{reflectivity_coeff}), the reflectivity coefficient for the point reflector located at $\check{\br}_l$, corresponding to the $(i,j)^{\rm th}$ pixel in the ROI, can be estimated as {
\begin{align}
\rho(\check{\br}_l) = \frac{s_l}{\displaystyle \int \limits_{-\infty}^{+\infty}g_1(\omega; \tilde{\br}_p, \hat{\br}) f_{12}(\omega;\hat{\br},\check{\br}_l,v)g_2(\omega;\hat{\br},\check{\br}_l,v)\,d\hat{x}} \label{21ee}
\end{align}}
where $s_l$ is the $l^{\rm th}$ entry of $\bs$. The uniqueness of the solution  of (\ref{l1_SMV}) is guaranteed if the spark of $\mathbf{\Phi}(\omega;v )$ is greater than $2L$ \cite{Zijan_Kruskal}, \cite[Theorem 2.4]{Chen_Kruskal}. The matrix $\mathbf{\Phi}(\omega;v)$ is a Vandermonde matrix and its spark is $M+1$. Therefore the uniqueness of the solution of (\ref{l1_SMV}) is guaranteed if $M > 2L$. However, in the presence of noise, the sparsity is inexact. Therefore, we need to guarantee the stability of the solution as well. It has been shown in \cite{DoBa} that if we {have} $M = O(L\;\log N)$, then stability is also guaranteed in the sense that the error is bounded.

{One of the reasons to use the $\ell_1$-norm minimization based approach instead of the  OGA is its stability. In fact the  optimization problem represented in (\ref{l1_SMV}) is a  global optimization problem. However, the stability of  the OGA  is local, which means that the stability is only valid for sufficiently small values of $\|w_p(\omega)\|_2$ \cite{Elad}.
Moreover, for moderately large $k$, where $k$ is the number of nonzero elements of the vector $\bs$, the $\ell_1$-norm minimization based technique can be very faster than OGA \cite{Foucart}.
In \cite{Elad,Donoho,Donoho1} the superiority of the $\ell_1$-norm minimization approach over the greedy based techniques have been  discussed at length.}\label{l1_OGA}

{
\paragraph{$\ell_1$-norm Minimization Solver for the SIMO Case}
In order to solve the $\ell_1$-norm minimization problem given in (\ref{l1_SMV}) we have used CVX \cite{cvx}, which is a package for specifying and solving convex programs. CVX supports five solvers: SeDuMi, SDPT3, Gurobi, MOSEK and GLPK. Each solver has different capabilities and different levels of performance. For instance, SeDuMi, SDPT3, and MOSEK 7 support all of the continuous (non-integer) models that CVX itself supports, while Gurobi is more limited, in that it does not support semidefinite constraints; and GLPK is limited even further. On the other hand, Gurobi, GLPK, and MOSEK support integer consraints, while SeDuMi and SDPT3 do not \cite{cvx_guide}.
SeDuMi and SDPT3 are included with the standard CVX distribution.
We have used CVX version 1.22. The solver by default is SeDuMi. SeDuMi is an open source interior point solver written in MATLAB. SeDuMi solver uses a variant of the primal-dual interior point method, which is known as the centering-predictor-corrector method \cite{sturm}.
SeDuMi solver of the CVX software package casts the optimization problem given in (\ref{l1_SMV}) as SOCP which using the interior point implementation the cost is $O(N^3)$ \cite{Lobo,Malioutov_journal}.}\label{cvx_SIMO}

{Below we analyse the error for the SIMO case when the  OGA is used to recover the sparse signal $\bs$.
\paragraph{Error Analysis for  the OGA in the SIMO Case}
Let assume that those columns of the matrix ${\mathbf{\Phi}}(\omega;v)$ corresponding to the nonzero elements of vector $\bs$ have been gathered in the sub-matrix ${\mathbf{\Phi}_s}(\omega;v)  \in \mathbb{C}^{M\times k}$. As a result we can rewrite (\ref{model_sparse_smv}) as
\begin{align}
\label{model_sparse_smv_prim}
\by_p(\omega) = [{\mathbf{\Phi}_s}(\omega;v) \; {\mathbf{\Phi}_{ns}}(\omega;v) ]\bu + \bw_p(\omega)
\end{align}
where we have partitioned the  matrix ${\mathbf{\Phi}}(\omega;v)$ into two sub-matrices ${\mathbf{\Phi}_s}(\omega;v)  \in \mathbb{C}^{M\times k}$ and ${\mathbf{\Phi}_{ns}}(\omega;v) \in \mathbb{C}^{M\times (N-k)}$. The scaler $k$ is the number of nonzero elements of $\bs$, i.e., $k = \|\bs\|_0$.  The matrix  ${\mathbf{\Phi}_s}(\omega;v)$ contains those $k$ columns of the matrix  ${\mathbf{\Phi}}(\omega;v)$ corresponding to the nonzero elements of the vector $\bs$ and the matrix  ${\mathbf{\Phi}_{ns}}(\omega;v)$ contains the rest of the columns of the matrix  ${\mathbf{\Phi}}(\omega;v)$. Moreover, the vector $\bu  \in \mathbb{R}^{M\times 1}$ is the same vector as the vector $\bs$ and the only difference is that the $k$ nonzero elements of the vector $\bs$ have been located in $u_1,u_2,\cdots,u_k$.  The result of the OGA, for the model described in (\ref{model_sparse_smv_prim}), is given as \cite{Elad}
\begin{align}
\label{OGA_noisy_smv}
{\bu}^{\flat} = {\mathbf{\Phi}^{-1}_s}(\omega;v )\by_p(\omega)
\end{align}
where the vector $\bu^{\flat}\in \mathbb{R}^{k\times 1}$ contains the $k$ nonzero elements of the vector $\bu$ and the matrix ${\mathbf{\Phi}^{-1}_s(\omega;v )}$ is the pseudo inverse of the matrix ${\mathbf{\Phi}_s}(\omega;v )$.
We rewrite (\ref{model_sparse_smv_prim}) for the noiseless case as
\begin{align}
\label{model_sparse_smv_prim_noiseless}
\by_p(\omega) = [{\mathbf{\Phi}_s}(\omega;v) \; {\mathbf{\Phi}_{ns}}(\omega;v) ]\bu_0
\end{align}
where the vector $\bu_0\in \mathbb{R}^{N\times 1}$ is the same vector as the vector $\bu$ but for the noiseless measurements.
The result of  the OGA for the noiseless model given in (\ref{model_sparse_smv_prim_noiseless}) is expressed as \cite{Elad}
\begin{align}
\label{OGA_noisless_smv}
{\bu}_0^{\flat} = {\mathbf{\Phi}^{-1}_s}(\omega;v )\by_p(\omega)
\end{align}
where the vector $\bu_0^{\flat}\in \mathbb{R}^{k\times 1}$ contains the $k$ nonzero elements of the vector $\bu_0$.
We manage to find the maximum error due to the presence of the noise. To accomplish this goal, we use (\ref{OGA_noisy_smv}) and (\ref{OGA_noisless_smv}) and write the following
\begin{align}
\label{error_smv_1}
 \| \bu^{\flat} -  \bu_0^{\flat}\|_2 & \leq \|{\mathbf{\Phi}^{-1}_s}(\omega;v )\bw_p(\omega) \|_2 \nonumber \\ & \leq  \|{\mathbf{\Phi}^{-1}_s}(\omega;v )\|_2\|\bw_p(\omega) \|_2
\end{align}
For the right hand side of (\ref{error_smv_1}) we have the following \cite{Matrix}
\begin{align}
\label{singular_value}
\|{\mathbf{\Phi}^{-1}_s}(\omega;v )\|_2\|\bw_p(\omega) \|_2 \leq \frac{\|\bw_p(\omega) \|_2}{\sigma_{\min}}
\end{align}
where $\sigma_{\min}$ is the minimum singular value of the matrix ${\mathbf{\Phi}_s}(\omega;v )$ and can be described as \cite{Matrix}
\begin{align}
\label{singular_value_1}
\sigma_{\min}=\|{\mathbf{\Phi}_s}(\omega;v )\|_2.
\end{align}
Hence, using (\ref{singular_value}) and (\ref{singular_value_1}) we can rewrite (\ref{error_smv_1}) as
\begin{align}
\label{error_smv_2}
 {\| \bu^{\flat} -  \bu^{\flat}_0\|_2} & \leq   \frac{\|\bw_p(\omega) \|_2}{\|{\mathbf{\Phi}_s}(\omega;v )\|_2}.
\end{align}
We also know that the following inequality holds \cite{mitzenmacher}
\begin{align}
\label{norm_ineq_1}
\frac{1}{\sqrt{k}} {\|{\mathbf{\Phi}_s}(\omega;v )\|_\infty} \leq  {\|{\mathbf{\Phi}_s}(\omega;v )\|_2} \leq  \sqrt{M}{\|{\mathbf{\Phi}_s}(\omega;v )\|_\infty}
\end{align}
where \cite{Matrix}
\begin{align}
\label{matrix_inf_norm}
\|{\mathbf{\Phi}_s}(\omega;v )\|_\infty = \max_{1\leq p \leq M} \sum\limits_{q = 1}^k |{\mathbf{\Phi}_s[p,q]}(\omega;v )|
\end{align}
where ${\mathbf{\Phi}_s[p,q]}(\omega;v )$ stands for the $(pq)^{\rm th}$ element of the matrix ${\mathbf{\Phi}_s}(\omega;v )$. Consequently from (\ref{norm_ineq_1}) we obtain
\begin{align}
\label{norm_ineq_2}
{\|{\mathbf{\Phi}_s}(\omega;v )\|_2} \geq \frac{1}{\sqrt{k}}{\|{\mathbf{\Phi}_s}(\omega;v )\|_\infty}.
\end{align}
We also have
\begin{align}
\label{norm_ineq_3}
{\|{\mathbf{\Phi}_s}(\omega;v )\|_\infty} \geq {\|{\mathbf{\Phi}_s}(\omega;v )\|_{\max}}
\end{align}
where \cite{Matrix}
\begin{align}
\label{Matrix_max_norm}
{\|{\mathbf{\Phi}_s}(\omega;v )\|_{\max}} = \max_{1\leq p \leq k, 1\leq q \leq M}|{\mathbf{\Phi}_s[p,q]}(\omega;v )|.
\end{align}
Therefore, using (\ref{Matrix_max_norm}), the relation given in (\ref{norm_ineq_2}) can be written as
\begin{align}
\label{norm_ineq_4}
{\|{\mathbf{\Phi}_s}(\omega;v )\|_2} \geq \frac{1}{\sqrt{k}}{\|{\mathbf{\Phi}_s}(\omega;v )\|_{\max}}.
\end{align}
Thus, using (\ref{norm_ineq_4}), we can rewrite (\ref{error_smv_2}) as
\begin{align}
\label{error_4}
  \| \bu^{\flat} - \bu^{\flat}_0\|_{2} \leq \sqrt{k}\frac{\|\bw_p(\omega)\|_2}{{\|{\mathbf{\Phi}_s}(\omega;v )\|_{\max}}}.
\end{align}
According to (\ref{steering_vector_element}), the $({ml})^{\rm th}$ element of the matrix  ${\mathbf{\Phi}_s}(\omega;v )$ for a hypothetical reflector located at $\check{\br}_l$ is given as $a_m(\omega; \check{\br}_l ,v)$ which after expansion is expressed as
\begin{align}
\label{ele_max_1}
& a_m(\omega; \check{\br}_l ,v) \triangleq \int \limits_{-\infty}^{+\infty}g_2(\omega;\hat{\br}, \check{\br}_l ,v)f_{21}(\omega;\tilde{\br}_m, \hat{\br})  g_1(\omega; \tilde{\br}_m, \hat{\br})\,d\hat{x}
 \nonumber \\ & = \int \limits_{-\infty}^{+\infty} \frac{e^{\displaystyle -j\omega \frac{\|\hat{\br} - \check{\br}_l\|}{v}}}{\|\hat{\br} - \check{\br}_l\|^{0.5}}\frac{j\omega |{\hat{z}}- {\tilde{z}}_m|}{4\pi c \|\tilde{\br}_m - \hat{\br}\|}\frac{e^{\displaystyle -j\omega \frac{\|\br_m - \hat{\br}\|}{c}}}{\|\br_m - \hat{\br}\|^{0.5}}\,d\hat{x}.
\end{align}
Consequently
\begin{align}
\label{ele_max_2}
|a_m(\omega; & \check{\br}_l ,v)| = \nonumber \\ & \int \limits_{-\infty}^{+\infty}  \frac{1}{\|\hat{\br} - \check{\br}_l\|^{0.5}}\frac{\omega |{\hat{z}}- {\tilde{z}}_m|}{4\pi c \|\tilde{\br}_m - \hat{\br}\|}\frac{1}{\|\br_m - \hat{\br}\|^{0.5}}\,d\hat{x}.
\end{align}
As we mentioned before we choose the first element  of the array from left as the origin of the coordinate system. Then the $\tilde{z}_m$ for the $m^{\rm th}$ receiver will be zero. Moreover, $\hat{z}$ will be the depth of the water above the test sample and is constant. Therefore, we can simplify (\ref{ele_max_2}) as
\begin{align}
\label{ele_max_3}
|a_m(\omega;&  \check{\br}_l ,v)|  =  \nonumber \\ & \frac{\omega|{\hat{z}}|}{4\pi c} \int \limits_{-\infty}^{+\infty}  \frac{1}{\|\hat{\br} - \check{\br}_l\|^{0.5}}\frac{1}{ \|\tilde{\br}_m - \hat{\br}\|}\frac{1}{\|\tilde{\br}_m - \hat{\br}\|^{0.5}}\,d\hat{x}.
\end{align}
Assume that for a specific indices $\hat{m}$ and $\hat{l}$, we have $|a_{\hat{m}}(\omega; \check{\br}_{\hat{l}} ,v)|\leq |a_m(\omega; \check{\br}_l ,v)|$ for all the possible values of $m$ and $r$. Therefore, we obtain\footnote{$
{\|{\mathbf{\Phi}_s}(\omega;v )\|_{\max}} = \max_{1\leq p \leq k, 1\leq q \leq M}|{\mathbf{\Phi}_s[p,q]}(\omega;v )|$ \cite{Matrix}.}
\begin{align}
\label{ele_max_phi}
& {\|{\mathbf{\Phi}_s}(\omega;v )\|_{\max}} \geq |a_{\hat{m}}(\omega; \check{\br}_{\hat{l}} ,v)|\nonumber \\
& = \frac{\omega|{\hat{z}}|}{4\pi c} \int \limits_{-\infty}^{+\infty} \frac{1}{\|\hat{\br} - \check{\br}_{\hat{l}}\|^{0.5}}\frac{1}{ \|\tilde{\br}_{\hat{m}} - \hat{\br}\|}\frac{1}{\|\tilde{\br}_{\hat{m}} - \hat{\br}\|^{0.5}}\,d\hat{x}.
\end{align}
Hence  using (\ref{ele_max_phi}) the error bound given in (\ref{error_4}) can be described  as
\begin{align}
\label{error_f}
 & \| \bu^{\flat} - \bu^{\flat}_0\|_2 \leq \nonumber \\& \sqrt{k} \frac{ \|\bw_p(\omega)\|_2}{\displaystyle \frac{\omega|{\hat{z}}|}{4\pi c} \int \limits_{-\infty}^{+\infty} \frac{1}{\|\hat{\br} - \check{\br}_{\hat{l}}\|^{0.5}}\frac{1}{ \|\tilde{\br}_{\hat{m}} - \hat{\br}\|}\frac{1}{\|\tilde{\br}_{\hat{m}} - \hat{\br}\|^{0.5}}\,d\hat{x}}.
\end{align}
The only term in (\ref{error_f}) that remains to be bounded is $\|\bw_p(\omega)\|_2$. To bound $\|\bw_p(\omega)\|_2$ we begin by the following lemma

\textbf{Lemma 1}: If $f :\; \mathbb{R}^n \rightarrow \mathbb{R}$ is $\lambda$-Lipschitz\footnote{A function $g:\; \mathbb{R}^n \rightarrow \mathbb{R}^m$ is said to be $\lambda$-Lipschitz, $\lambda\geq 0$, if $|g(\ba)-g(\bb)|\leq \lambda |\ba-\bb|$, for every $\ba,\bb \in \mathbb{R}^n$ \cite{Rudin}.}, then $ \forall\epsilon>0$
\begin{align}
\label{lemma}
\gamma_n(\bu\in\mathbb{R}^n:|f(\bw_p(\omega))-\mathbb{E} (f(\bw_p(\omega)))|\geq \epsilon) \leq & e^{\frac{-\epsilon^2}{2\lambda^2}}
\end{align}
where $\gamma_n$ is the Gaussian measure in $n$ dimension.

\textbf{Proof}. See \cite{Ledoux,Milman, Measure_concentration}.

The function, $f(\cdot)$, is $\|(\cdot)\|_2 $ which is 1-Lipschitz. Therefore, $\lambda = 1$. Furthermore, since each $w_{pi}(\omega)\thicksim \mathcal{N}(0,\sigma^2)$  for $i \in \{ 1,2,\cdots,M\}$, therefore the expression  $\mathbb{E} (f(\bw_p(\omega))) = \mathbb{E} (\|\bw_p(\omega)\|_2)$ can be bounded as
\begin{align}
\label{w_ineq}
\mathbb{E} (\|\bw_p(\omega)\|_2) =  \mathbb{E} {\left\{\sum\limits_{i=1}^{M}|w_i|^2\right\}}^{\frac{1}{2}}&  \leq  {\left\{\sum\limits_{i=1}^{M}\mathbb{E}(|w_i|^2)\right\}}^{\frac{1}{2}}  \nonumber \\ & = \sqrt{M}\sigma
\end{align}
where we have used Jensen's inequality for the concave functions \cite{mitzenmacher}. Hence, from (\ref{lemma}) we obtain
\begin{align}
\label{w_bound}
p(|\|\bw_p(\omega)\|_2-\mathbb{E} (\|\bw_p(\omega)\|_2)|\geq \epsilon) \leq e^{\frac{-\epsilon^2}{2}}, \;\;\; \forall\epsilon>0.
\end{align}
According to (\ref{w_bound}),  $\|\bw_p(\omega)\|_2$ concentrates around its mean with high probability. Thus, using the upper bound obtained in (\ref{w_ineq})  for $\mathbb{E} (\|\bw_p(\omega)\|_2)$, we obtain that with high probability
\begin{align}
\label{w_bound_2}
\|\bw_p(\omega)\|_2\leq\sqrt{M}\sigma.
\end{align}
Hence, using (\ref{w_bound_2}), the error bound given in (\ref{error_f}) is described as
\begin{align}
\label{error_smv}
  &\| \bu^{\flat} - \bu^{\flat}_0\|_{2} \leq \nonumber \\ & \sqrt{kM} \frac{\sigma }{\displaystyle \frac{\omega|{\hat{z}}|}{4\pi c} \int \limits_{-\infty}^{+\infty} \frac{1}{\|\hat{\br} - \check{\br}_{\hat{l}}\|^{0.5}}\times\frac{1}{ \|\tilde{\br}_{\hat{m}} - \hat{\br}\|}\times\frac{1}{\|\tilde{\br}_{\hat{m}} - \hat{\br}\|^{0.5}}\,d\hat{x}}.
\end{align}
The error bound for $\bu$ is exactly the same as the error bound given for $\bu^{\flat}$ in (\ref{error_smv}).}\label{error_OGA_SIMO}

{
\paragraph{Error Analysis for the $\ell_1$-norm Minimization Based Approach for the SIMO Case}
For the optimization problem given in (\ref{l1_SMV})  we have \cite[Theorem 4.19]{Foucart}
\begin{align}
\label{l1_ssv_error_1}
\|\bs - \bs_0\|_1 \leq \frac{2+2\rho_s}{1-\rho_s}\sigma_s(\bs_0)_1 + \frac{4\tau_s}{1-\rho_s}\|\bw_p(\omega)\|_2
\end{align}
where $\rho_s$ is a constant such that $0<\rho_s < 1$, $\tau_s>0$. The term $\sigma_k(\bs_0)_1$ is the $\ell_1$-error of the best $k$-term approximation to a vector $\bs_0$ and is defined as
\begin{align}
\label{sigma_k_1}
\sigma_{\rm k}(\bs_0)_1 \triangleq \inf \{\|\bs_0-\bz\|_1, \; \bz \in \mathbb{C}^{N\times 1}\;{\rm is \; k\;sparse}\}.
\end{align}
We also have the following \cite{Foucart}
\begin{align}
\label{sigma_k_2}
\sigma_{\rm k}(\bs_0)_2 \leq \frac{1}{2\sqrt{\rm k}}\|\bs_0\|_1.
\end{align}
Since $\|\bs_0-\bz\|_1 \leq \sqrt{N}\|\bs_0-\bz\|_2$ then based on (\ref{sigma_k_1}), we can write $\sigma_{\rm k}(\bs_0)_1 \leq \sqrt{N} \sigma_{\rm k}(\bs_0)_2$. Therefore, we can rewrite (\ref{l1_ssv_error_1}) as
\begin{align}
\label{l1_ssv_error_2}
\|\bs - \bs_0\|_1 \leq \frac{2+2\rho_s}{1-\rho_s}\sqrt{N} \sigma_{\rm k}(\bs_0)_2 + \frac{4\tau_s}{1-\rho_s}\sqrt{M}\sigma
\end{align}
where for the second part of the right hand side of (\ref{l1_ssv_error_1}) we have used (\ref{w_bound_2}), i.e.,  $\|\bw_p(\omega)\|_2\leq\sqrt{M}\sigma$.
Using (\ref{sigma_k_2}) we can describe (\ref{l1_ssv_error_2}) as
\begin{align}
\label{l1_ssv_error_3}
\|\bs - \bs_0\|_1 \leq \frac{(1+\rho_s)}{(1-\rho_s)\sqrt{\rm k}} \sqrt{N} \|\bs_0\|_1 + \frac{4\tau_s}{1-\rho_s}\sqrt{M}\sigma.
\end{align}
Using $\|\bs_0\|_1 \leq N \|\bs_0\|_{\infty}$ we can express (\ref{l1_ssv_error_3}) as
\begin{align}
\label{l1_ssv_error_4}
\|\bs - \bs_0\|_{1} \leq \frac{(1+\rho_s)}{(1-\rho_s)\sqrt{\rm k}} \sqrt{N} N \|\bs_0\|_{\infty} + \frac{4\tau_s}{1-\rho_s}\sqrt{M}\sigma
\end{align}
Let $\boldsymbol{\delta} \in \mathcal{N}$, where  $\mathcal{N}$ is the null space of the matrix ${\mathbf{\Phi}}(\omega ;v)$. We further define the set $I$ to be a set of the indices of the nonzero elements of $\bu_0$, i.e., $I={\rm supp}(\bu_0)$, where $\bu_0$ has been given in (\ref{model_sparse_smv_prim_noiseless}).  Then we partition $\boldsymbol{\delta}$ as $\boldsymbol{\delta}=(\boldsymbol{\delta}_I,\boldsymbol{\delta}_{I^c})$, where $I^c$ is the complement of $I$. The matrix ${\mathbf{\Phi}}(\omega ;v)$ is said to satisfy the stable  null space property with constant $0<\rho_s < 1$ if \cite{Foucart,Donoho,Donoho1}
\begin{align}
\label{null-space_property_1}
\|\boldsymbol{\delta}_I \|_1\leq \rho_s \|\boldsymbol{\delta}_{I^c}\|_1.
\end{align}
We manage to find $\rho_s$. To accomplish this goal we define $\bv={\mathbf{\Phi}}_s(\omega ;v)I = {\mathbf{\Phi}}_{ns}(\omega ;v)I^c $. Then we can write the following
\begin{align}
\label{null-space_property_2}
\|\boldsymbol{\delta}_I\|_1 \leq \sqrt{k}\|\boldsymbol{\delta}_{I}\|_2.
\end{align}
We can further write
\begin{align}
\label{null-space_property_3}
\|\bv\|_2 = \|{\mathbf{\Phi}}_s(\omega ;v)\boldsymbol{\delta}_{I}\|_2 \geq \sigma_{{\min}_s} \|\boldsymbol{\delta}_{I}\|_2
\end{align}
where $\sigma_{{\min}_s}$ stands for the minimum singular value of the matrix ${\mathbf{\Phi}}_s(\omega ;v)$. We also know that $\sigma_{{\min}_s} =\|{\mathbf{\Phi}}_s(\omega ;v)\|_2$, hence using (\ref{null-space_property_3}) the relation (\ref{null-space_property_2}) is given as
\begin{align}
\label{null-space_property_4}
\|\boldsymbol{\delta}_I\|_1 \leq \sqrt{k} \frac{\|\bv\|_2}{\|{\mathbf{\Phi}}_s(\omega ;v)\|_2} \leq \sqrt{k} \frac{\|\bv\|_1}{\|{\mathbf{\Phi}}_s(\omega ;v)\|_2}.
\end{align}
To connect $\|\bv\|_1$ to $\|\boldsymbol{\delta}_{{I}^c}\|_1$ we use $\bv={\mathbf{\Phi}}_s(\omega ;v)\boldsymbol{\delta}_I = {\mathbf{\Phi}}_{ns}(\omega ;v)\boldsymbol{\delta}_{{I}^c}$ which upon taking norm-1 from both sides we obtain
\begin{align}
\label{null-space_property_5}
\|\bv\|_1= \|{\mathbf{\Phi}}_{ns}(\omega ;v)\boldsymbol{\delta}_{I^c}\|_1 \leq \|{\mathbf{\Phi}}_{ns}(\omega ;v)\|_1\|\boldsymbol{\delta}_{I^c}\|_1.
\end{align}
Using (\ref{null-space_property_5}) the modified version of (\ref{null-space_property_4}) can be described as
\begin{align}
\label{null-space_property_6}
\|\boldsymbol{\delta}_I\|_1 \leq \sqrt{k} \frac{\|{\mathbf{\Phi}}_{ns}(\omega ;v)\|_1}{\|{\mathbf{\Phi}}_s(\omega ;v)\|_2}\|\boldsymbol{\delta}_{{I}^c}\|_1.
\end{align}
Therefore, based on (\ref{null-space_property_6}), the parameter $\rho$ in (\ref{null-space_property_1}) is given as
\begin{align}
\label{rho_1}
 \rho_s = \sqrt{k} \frac{\|{\mathbf{\Phi}}_{ns}(\omega ;v)\|_1}{\|{\mathbf{\Phi}}_s(\omega ;v)\|_2}.
\end{align}
Moreover, instead of $\bs$ and $\bs_0$ we can work with $\bs^{\prime}=\displaystyle \frac{\bs}{\max\{\bs\}}$ and $\bs^{\prime}_0= \displaystyle \frac{\bs_0}{\max\{\bs_0\}}$. As a result $\|\bs^{\prime}_0\|_{\infty}=1$ and we can rewrite (\ref{l1_ssv_error_4}) as
\begin{align}
\label{l1_ssv_error_5}
\|\bs^{\prime} - \bs_0^{\prime}\|_{1} \leq \frac{(1+\rho_s)}{(1-\rho_s)\sqrt{\rm k}} \sqrt{N} N  + \frac{4\tau_s}{1-\rho_s}\sqrt{M}\sigma.
\end{align}
Consequently, using (\ref{rho_1}) the error bound given in (\ref{l1_ssv_error_5}) can be described as
\begin{align}
\label{l1_ssv_error_6}
 \|\bs^{\prime} - \bs_0^{\prime}\|_{1}&  \leq  \frac{\left(\|{\mathbf{\Phi}}_s(\omega ;v)\|_2+\sqrt{k}\|{\mathbf{\Phi}}_{ns}(\omega ;v)\|_1\right) \sqrt{N/k} N}{\left(\|{\mathbf{\Phi}}_s(\omega ;v)\|_2-\sqrt{k}\|{\mathbf{\Phi}}_{ns}(\omega ;v)\|_1\right)} \nonumber \\ & +      \frac{4\tau_s\|{\mathbf{\Phi}}_s(\omega ;v)\|_2\sqrt{M}\sigma}{\left(\|{\mathbf{\Phi}}_s(\omega ;v)\|_2-\sqrt{k}\|{\mathbf{\Phi}}_{ns}(\omega ;v)\|_1\right)}.
\end{align}}\label{error_l1_SIMO}

Image reconstruction procedure based on (\ref{l1_SMV}) and \eqref{21ee} utilizes only one snapshot. Although utilizing one snapshot is one of the advantages of this technique, since we have access to multiple snapshots, we can use all these snapshots to improve the quality of the resulting image.

\subsection{Multiple Input Multiple Output Case}
In this section, we present the generalization of (\ref{l1_SMV}) for the MIMO case. To develop the model for the MIMO case, we define
\begin{align}
\label{steering vector_p}
\ba_p(\omega; \br ,v) \triangleq [ a_{1p}(\omega; \br , v)\;a_{2p}(\omega; \br ,v)\;\cdots\;a_{Mp}(\omega; \br ,v)]^T.
\end{align}
In (\ref{steering vector_p}), $a_{mp}(\omega; \br ,v)$, for $m \in \{1,\cdots,M\}$, is described as
\begin{align}
\label{array_manifold_component_p}
a_{mp}(& \omega; \br ,v)  \triangleq \int \limits_{-\infty}^{+\infty}g_2(\omega;\hat{\br}, \br ,v)f_{21}(\omega;\tilde{\br}_m, \hat{\br})  g_1(\omega; \tilde{\br}_m, \hat{\br})\,d\hat{x} \nonumber \\
& \times \int \limits_{-\infty}^{+\infty}[g_1(\omega; \tilde{\br}_p, \hat{\br}) f_{12}(\omega;\hat{\br},\br ,v)g_2(\omega;\hat{\br},\br ,v)]\,d\hat{x}.
\end{align}
In fact, (\ref{steering vector_p}) contains the signal propagation effect for both the transmission and the reception paths. However, the signal propagation effect for transmission case depends on the transducer's number. Therefore, for each transmitting transducer, we have a different signature and that is the reason for the index $p$ in (\ref{steering vector_p}). The data model in (\ref{data}) is modified as {
\begin{align}
\by(\omega) = {\underline{\bA}(\omega )\underline{\boldsymbol{\rho}}}+ \bw(\omega )
\end{align}}
where the vector $\by(\omega)\in \mathbb{C}^{M^2\times 1}$ is described as
\begin{align}
\label{all_measurements}
\by(\omega)  = [\by^T_1(\omega)\;\by^T_2(\omega)\;\cdots\;\by^T_M(\omega)]^T
\end{align}
and the matrix $\underline{\bA}(\omega )\in \mathbb{C}^{M^2\times L}$ is defined as
\begin{align}
\label{over_complete}
\underline{\bA}(\omega ) \triangleq [ \underline{\bA}^T_1(\omega ) \;  \underline{\bA}^T_2(\omega )\;\cdots\; \underline{\bA}^T_M(\omega )]^T.
\end{align}
Here, the $l^{\rm th}$ column of the matrix $\underline{\bA}_p(\omega)\in \mathbb{C}^{M\times L}$ is the array steering vector $\ba_p(\omega; \check{\br}_l ,v) \in \mathbb{C}^{M\times 1}$,  given in (\ref{steering vector_p}), when $\br$ is replaced with $\breve \br_l$ . Also, the vector { $\underline{\boldsymbol{\rho}}\in \mathbb{R}^{L\times 1}$} is defined as {
\begin{align}
\label{ref_coeff_mmv}
\underline{\boldsymbol{\rho}} = [\rho(\check{r}_1) \;\rho(\check{r}_2) \;\; \cdots \;\rho(\check{r}_L)]^T.
\end{align}}
Finally, the noise vector $\bw(\omega)\in \mathbb{C}^{M^2\times 1}$ is described as
\begin{align}
\bw(\omega) = [ \bw^T_1(\omega)\; \bw^T_2(\omega)\;\cdots\; \bw^T_M(\omega)]^T
\end{align}
where the vector $\bw_p(\omega) \in \mathbb{C}^{M\times 1}$ is the noise received by the array when the $p^{\rm th}$ transducer transmits.
Then a new dictionary based on the steering vector given in (\ref{steering vector_p}), for the case when the $p^{\rm th}$ transducer transmits, is defined as
\begin{align}
\label{phi_p}
{\mathbf{\Phi}}_p(\omega;v) \triangleq [\ba_p(\omega; \check{\br}_1 ,v)\; \ba_p(\omega; \check{\br}_2 ,v)\; \cdots  \;\ba_p(\omega; \check{\br}_N,v)].
\end{align}
{Therefore}, the sparse signal representation based model for the MIMO case, is described as
\begin{align}
\label{model_mmv}
\by(\omega) =  \tilde{\mathbf{\Phi}}(\omega ;v) \underline{\bs} + \bw(\omega)
\end{align}
where the $l^{\rm th}$ element of $\underline{\bs} \in \mathbb{C}^{N\times 1}$ is nonzero if there is a scatterer located at the $(i,j)^{\rm th}$ pixel, where $i= \lfloor\frac{l-1}{n_x}\rfloor+1$ and $ j=l-(i-1)n_x $, and it is zero otherwise. Each nonzero element of $\underline{\bs}$ is one of the entries of $\underline{\boldsymbol{\rho}}$ in (\ref{ref_coeff_mmv}). The matrix $\tilde{\mathbf{\Phi}}(\omega ;v) \in \mathbb{C}^{M^2\times N}$ {in (\ref{model_mmv})} is defined as
\begin{align}
\label{phi_mmv}
\tilde{\mathbf{\Phi}}(\omega ;v)  \triangleq [\mathbf{\Phi}^T_1(\omega ;v) \;\mathbf{\Phi}^T_2(\omega ;v)\;\cdots\;\mathbf{\Phi}^T_M(\omega ;v)]^T.
\end{align}
Here, $\mathbf{\Phi}_p(\omega ;v) \in \mathbb{C}^{M\times N}$ is the dictionary given in (\ref{phi_p}). The optimization problem for the MIMO case is then expressed as {
\begin{align}
 \label{l1_mmv}
 \rm{\min_{\underline{\bs}}}. \ & \|\underline{\bs}\|_1 \nonumber \\
 \rm{subject \ to} \ & \|\by(\omega) - \tilde{\mathbf{\Phi}}(\omega ;v) \underline{\bs}\|_2\le \beta_m \nonumber \\
 & \underline{\bs} \succeq 0, \;\;\; \underline{\bs}\in \mathbb{R}_{+}^{N \times 1}
 \end{align}}
{where $\beta_m$ is the regularization parameter and is chosen such that the probability of the norm of the noise vector being larger than $\beta_m$ is small \cite{Malioutov_journal,Thesis}. To solve (\ref{l1_mmv}), which is a SOCP optimization problem, we can use any software packages dedicated to solving convex optimization problems such as CVX software package \cite{cvx}. The CVX software package can easily solve (\ref{l1_mmv}) even for large scale problems \cite{Boyd_book}.}\label{beta_m}
 By reshaping $ \underline{\bs}$ into an $n_z \times n_x$ matrix, we obtain the final image. The uniqueness of the solution to (\ref{l1_mmv}) is guaranteed, if the spark of $\tilde{\mathbf{\Phi}}(\omega ;v)$ which is $M^2+1$, satisfies {$M^2 > (2L + 1 - \rm rank(\by(\omega)))$}  \cite{Zijan_Kruskal}, \cite[Theorem 2.4]{Chen_Kruskal}. This shows that the upper limit on $L$ has been improved by $\rm rank(\by(\omega))$ comparing to the upper limit that we had for SIMO case.  {Based on (\ref{all_measurements}), $\rm rank(\by(\omega)) = 1$. Hence the uniqueness of the solution to (\ref{l1_mmv}) is guaranteed if $M^2 > 2L$}. For the stability of the result we need to choose {$M^2 = O(L\;\rm{log}N)$}, which guarantees that the error is bounded\cite{DoBa}. Of course we must still ensure that {$N > M^2$} holds true.

{
\paragraph{$\ell_1$-norm Minimization Solver for the MIMO Case}
The optimization problems given in (\ref{l1_mmv}) is in standard SOCP form.
To solve (\ref{l1_mmv}), SeDuMi  solver uses the barrier method or path following method \cite{Boyd_book}.  The inequalities in (\ref{l1_mmv}) define the positive orthant in $\mathbb{R}^{N\times 1}$. These can be equipped with the following self-concordant barrier (in fact these are $\nu$-self-concordant barriers)\footnote{Definition: The Standard self-concordant barrier $F(x)$ is called $\nu$-self-concordant barrier if \cite{Nesterov}
\begin{align}
\max_{u\in\mathbb{R}_{+}^{1}}[2<F^{\prime}(x),u>-<F^{\prime \prime }(x)u,u>]\leq\nu \nonumber\;\;\; \forall x \in {\rm dom}\;F
\end{align}}
\begin{align}
\label{barrier_mmv}
F_m(\underline{\bs}) = -\sum\limits_{i = 1}^N \log(\underline{\bs}(i)),  \;\;\; \nu_m=N.
\end{align}
This barrier is called the standard logarithmic barrier for $\mathbb{R}_{+}^{N\times 1}$. In fact the standard logarithmic barrier is optimal for $\mathbb{R}_{+}^{N\times 1}$ \cite[Theorem 4.3.2]{Nesterov}.
The complexity estimates of solving (\ref{l1_mmv}) using SeDuMi solver is $O(\sqrt{N}\ln(\frac{1}{\epsilon_m}))$ \cite{sturm}. The parameter $\epsilon_m$ is given as $|\underline{\bs} - \underline{\bs}^{*}| \leq \epsilon_m$, where  $\underline{\bs}^{*}$  stands for the optimal value of $\underline{\bs}$ in (\ref{l1_mmv}).}\label{cvx_MIMO}

{
\paragraph{Error Analysis for the OGA in the MIMO Case}
Let assume that those columns of the matrix $\tilde{\mathbf{\Phi}}(\omega ;v)$ corresponding to the nonzero elements of vector $ \underline{\bs}$ have been gathered in the sub-matrix $\tilde{\mathbf{\Phi}}_s(\omega ;v)\in \mathbb{C}^{M\times k}$. Therefore, we rewrite (\ref{model_mmv}) as
\begin{align}
\label{model_sparse_mmv}
\by(\omega) =  [\tilde{\mathbf{\Phi}}_s(\omega ;v) \;\tilde{\mathbf{\Phi}}_{sn}(\omega ;v)]\underline{\bu} + \bw(\omega)
\end{align}
where we have partitioned  matrix $\tilde{\mathbf{\Phi}}(\omega ;v)$ into two sub-matrices $\tilde{\mathbf{\Phi}}_s(\omega ;v)  \in \mathbb{C}^{M^2\times k}$ and $\tilde{\mathbf{\Phi}}_{sn}(\omega ;v) \in \mathbb{C}^{M^2\times (N-k)}$. The scaler $k$ is the number of nonzero elements of $ \underline{\bs}$, i.e., $k = \|\underline{s}\|_0$, and $\tilde{\mathbf{\Phi}}_{s}(\omega ;v)$ contains those $k$ columns of $\tilde{\mathbf{\Phi}}(\omega;v)$ corresponding to the nonzero elements of the vector $\underline{\bs}$ and $\tilde{\mathbf{\Phi}}_{sn}(\omega ;v)$ contains the rest of the columns of the matrix $\tilde{\mathbf{\Phi}}(\omega ;v)$.  Moreover, the vector $\underline{\bu}  \in \mathbb{R}^{N\times 1}$ is the same vector as the vector $\underline{\bs}$ and the only difference is that the $k$ nonzero elements of the vector $\underline{\bs}$ have been located in $\underline{u}_1,\underline{u}_2,\cdots,\underline{u}_k$. The result of the OGA, for the model described in (\ref{model_sparse_mmv}), is given as \cite{Elad}
\begin{align}
\label{OGA_noisy}
 \underline{\bu}^{\flat} = {\tilde{\mathbf{\Phi}}^{-1}_s}(\omega;v )\by(\omega)
\end{align}
where the vector $ \underline{\bu}^{\flat}\in \mathbb{R}^{k\times 1}$ contains the $k$ nonzero elements of the vector $ \underline{\bu}$ and the matrix ${\mathbf{\Phi}^{-1}_s(\omega;v )}$ is the pseudo inverse of the matrix $\tilde{\mathbf{\Phi}}_s(\omega;v )$.
We rewrite (\ref{model_sparse_mmv}) for the noiseless case as
\begin{align}
\label{model_sparse_mmv_noiseless}
\by(\omega) = [\tilde{\mathbf{\Phi}}_s(\omega;v) \; \tilde{\mathbf{\Phi}}_{ns}(\omega;v) ] \underline{\bu}_0
\end{align}
where the vector $ \underline{\bu}_0\in \mathbb{R}^{N\times 1}$ is the same vector as the vector $ \underline{\bu}$ but for the noiseless measurements.
The result of  the OGA for the noiseless model given in (\ref{model_sparse_mmv_noiseless}) is given as \cite{Elad}
\begin{align}
\label{OGA_noisless}
{ \underline{\bu}}_0^{\flat} = \tilde{\mathbf{\Phi}}^{-1}_s(\omega;v )\by(\omega)
\end{align}
where the vector $ \underline{\bu}_0^{\flat}\in \mathbb{R}^{k\times 1}$ contains the $k$ nonzero elements of the vector $ \underline{\bu}_0$.
We manage to find the maximum error due to the presence of the noise. To accomplish this goal, we use (\ref{OGA_noisy}) and (\ref{OGA_noisless}) and write the following
\begin{align}
\label{error_0}
 \| \bu^{\flat} -  \bu_0^{\flat}\|_2 & \leq \|{\mathbf{\Phi}^{-1}_s}(\omega;v )\bw_p(\omega) \|_2 \nonumber \\ &  \leq  \|{\mathbf{\Phi}^{-1}_s}(\omega;v )\|_2\|\bw_p(\omega) \|_2
\end{align}
An element of the matrix  ${\tilde{\mathbf{\Phi}}_s}(\omega;v )$ for a hypothetical reflector located at $\check{\br}_l$ is  $a_{mp}(\omega; \check{\br}_l ,v)$ as given in (\ref{array_manifold_component_p}). Following the same procedure that we did for the SIMO  case we obtain
\begin{align}
\label{ele_max_p_1}
& a_{mp}(\omega; \check{\br}_l ,v)  \triangleq \nonumber \\ & \int \limits_{-\infty}^{+\infty}g_2(\omega;\hat{\br}, \check{\br}_l ,v)f_{21}(\omega;\tilde{\br}_m, \hat{\br})  g_1(\omega; \tilde{\br}_m, \hat{\br})\,d\hat{x} \nonumber \\
&\times \int \limits_{-\infty}^{+\infty}[g_1(\omega; \tilde{\br}_p, \hat{\br}) f_{12}(\omega;\hat{\br},\check{\br}_l ,v)g_2(\omega;\hat{\br},\check{\br}_l ,v)]\,d\hat{x}
\end{align}
which yields
\begin{align}
\label{ele_max_p_1}
 |a_{mp}(& \omega; \check{\br}_l ,v)|  = \nonumber \\ & \frac{\omega|{\hat{z}}|}{4\pi c} \int \limits_{-\infty}^{+\infty} \frac{1}{\|\hat{\br} - \check{\br}_l\|^{0.5}}\frac{1}{ \|\tilde{\br}_m - \hat{\br}\|}\frac{1}{\|\tilde{\br}_m - \hat{\br}\|^{0.5}}\,d\hat{x} \nonumber \\
& \frac{\omega}{4\pi v}\int \limits_{-\infty}^{+\infty}\frac{1}{\|\tilde{\br}_p- \hat{\br}\|^{0.5}}\frac{ |z_l-\hat{z}|}{ \|\hat{\br} - \check{\br}_l\|} \frac{1}{\|\hat{\br}-\check{\br}_l\|^{0.5}}\,d\hat{x}.
\end{align}
Assume that for a specific indices $\hat{m}$, $\hat{p}$ and $\hat{l}$, we have $|a_{\hat{m}\hat{p}}(\omega; \check{\br}_{\hat{l}} ,v)|\leq |a_{mp}(\omega; \check{\br}_l ,v)|$ for all the possible values of $m$, $p$ and $r$. Therefore, we obtain\footnote{$
{\|{\mathbf{\Phi}_s}(\omega;v )\|_{\max}} = \max_{1\leq p \leq k, 1\leq q \leq M}|{\mathbf{\Phi}_s[p,q]}(\omega;v )|$ \cite{Matrix}.}
\begin{align}
\label{ele_max_phi_MIMO}
&{\|\tilde{\mathbf{\Phi}}_s(\omega ;v) \|_{\max}}  \geq |a_{\hat{m}\hat{p}}(\omega; \check{\br}_{\hat{l}} ,v)|
 \nonumber \\ & =  \frac{\omega|{\hat{z}}|}{4\pi c} \int \limits_{-\infty}^{+\infty} \frac{1}{\|\hat{\br} - \check{\br}_{\hat{l}}\|^{0.5}}\times\frac{1}{ \|\tilde{\br}_{\hat{m}} - \hat{\br}\|}\times\frac{1}{\|\tilde{\br}_{\hat{m}} - \hat{\br}\|^{0.5}}\,d\hat{x} \times\nonumber \\
& \frac{\omega}{4\pi v}\int \limits_{-\infty}^{+\infty}\frac{1}{\|\tilde{\br}_{\hat{p}}- \hat{\br}\|^{0.5}}\times \frac{ |z_l-\hat{z}|}{ \|\hat{\br} - \check{\br}_{\hat{l}}\|}\times \frac{1}{\|\hat{\br}-\check{\br}_{\hat{l}}\|^{0.5}}\,d\hat{x}.
\end{align}
Hence the error bound given in (\ref{error_f}), for the SIMO case, can be rewritten for the MIMO case as
\begin{align}
\label{error_p_f}
 & \| { \underline{\bu}^{\flat}} - \underline{\bu}^{\flat}_0\|_{2}  \leq \nonumber \\ & \sqrt{k}\frac{\|\bw(\omega)\|_2}{ \displaystyle \frac{\omega^2|{\hat{z}}|}{16\pi^2 cv} \int \limits_{-\infty}^{+\infty} \frac{1}{\|\hat{\br} - \check{\br}_{\hat{l}}\|^{0.5}}\times\frac{1}{ \|\tilde{\br}_{\hat{m}} - \hat{\br}\|}\times\frac{1}{\|\tilde{\br}_{\hat{m}} - \hat{\br}\|^{0.5}}\,d\hat{x}}\nonumber \\ & \times\frac{1}{ \displaystyle
\int \limits_{-\infty}^{+\infty}\frac{1}{\|\tilde{\br}_{\hat{p}}- \hat{\br}\|^{0.5}}\times \frac{ |z_l-\hat{z}|}{ \|\hat{\br} - \check{\br}_{\hat{l}}\|}\times \frac{1}{\|\hat{\br}-\check{\br}_{\hat{l}}\|^{0.5}}\,d\hat{x}}.
\end{align}
To bound $\|\bw(\omega)\|_2$ we use the same procedure as we did to bound $\|\bw_p(\omega)\|_2$. The only difference is the size of two vectors. Therefore, using (\ref{w_bound_2}) but this time for $\|\bw(\omega)\|_2$,
\begin{align}
\label{w_ineq_msv}
\mathbb{E} (\|\bw(\omega)\|_2) =  \mathbb{E} {\left\{\sum\limits_{i=1}^{M^2}|w_i|^2\right\}}^{\frac{1}{2}}  \leq  {\left\{\sum\limits_{i=1}^{M^2}\mathbb{E}(|w_i|^2)\right\}}^{\frac{1}{2}} = M\sigma
\end{align}
where we have used Jensen's inequality for the concave functions \cite{mitzenmacher}. Hence, from (\ref{lemma}) we obtain
\begin{align}
\label{w_bound_msv}
p(|\|\bw(\omega)\|_2-\mathbb{E} (\|\bw(\omega)\|_2)|\geq \epsilon) \leq e^{\frac{-\epsilon^2}{2}}, \;\;\;\;\forall \epsilon>0.
\end{align}
According to (\ref{w_bound_msv}),  $\|\bw(\omega)\|_2$ concentrates around its mean with high probability. Thus, using the upper bound given in (\ref{w_ineq_msv}) for $\mathbb{E} (\|\bw(\omega)\|_2)$, we obtain that with high probability
\begin{align}
\label{w_bound_3}
\|\bw(\omega)\|_2\leq M\sigma.
\end{align}
Hence using (\ref{w_bound_3}), the error bound in (\ref{error_p_f}) is expressed as
\begin{align}
\label{error_mmv}
 & \| { \underline{\bu}^{\flat}} - \underline{\bu}^{\flat}_0\|_{2}  \leq \nonumber \\ & \sqrt{k}\frac{M\sigma}{ \displaystyle \frac{\omega^2|{\hat{z}}|}{16\pi^2 cv} \int \limits_{-\infty}^{+\infty} \frac{1}{\|\hat{\br} - \check{\br}_{\hat{l}}\|^{0.5}}\times\frac{1}{ \|\tilde{\br}_{\hat{m}} - \hat{\br}\|}\times\frac{1}{\|\tilde{\br}_{\hat{m}} - \hat{\br}\|^{0.5}}\,d\hat{x}}\nonumber \\ & \times\frac{1}{ \displaystyle
\int \limits_{-\infty}^{+\infty}\frac{1}{\|\tilde{\br}_{\hat{p}}- \hat{\br}\|^{0.5}}\times \frac{ |z_l-\hat{z}|}{ \|\hat{\br} - \check{\br}_{\hat{l}}\|}\times \frac{1}{\|\hat{\br}-\check{\br}_{\hat{l}}\|^{0.5}}\,d\hat{x}}.
\end{align}
Finally the error bound for $ \underline{\bu}$ is exactly the same as the error bound given for $ \underline{\bu}^{\flat}$ in (\ref{error_mmv}).}\label{error_OGA_MIMO}

{
\paragraph{Error Analysis for the $\ell_1$-norm Minimization based Method for the MIMO Case}
For the optimization problem given in (\ref{l1_mmv}) we have \cite[Theorem 4.19]{Foucart}
\begin{align}
\label{l1_mmv_error_1}
\|\underline{\bs} - \underline{\bs}_0\|_1 \leq \frac{2+2\rho_m}{1-\rho_m}\sigma_m(\underline{\bs}_0)_1 + \frac{4\tau_m}{1-\rho_m} \|\bw(\omega)\|_2
\end{align}
where $\rho_m$ is a constant such that $0< \rho_m < 1$ and $\tau_m>0$. The term $\sigma_k(\underline{\bs}_0)_1$ is the $\ell_1$-error of the best $k$-term approximation to a vector $\underline{\bs}_0$ and is defined as \cite{Foucart}
\begin{align}
\label{sigma_mmv_k_1}
\sigma_{\rm k}(\underline{\bs}_0)_1 \triangleq \inf \{\|\underline{\bs}_0-\underline{\bz}\|_1, \; \underline{\bz} \in \mathbb{R}^{N\times 1}\;{\rm is \; k\;sparse}\}.
\end{align}
We also have the following \cite{Foucart}
\begin{align}
\label{sigma_mmv_k_2}
\sigma_{\rm k}(\underline{\bs}_0)_2 \leq \frac{1}{2\sqrt{\rm k}}\|\underline{\bs}_0\|_1.
\end{align}
Since $\|\underline{\bs}_0-\underline{\bz}\|_1 \leq \sqrt{N}\|\underline{\bs}_0-\underline{\bz}\|_2$ then based on (\ref{sigma_mmv_k_1}), we can write $\sigma_{\rm k}(\underline{\bs}_0)_1 \leq \sqrt{N} \sigma_{\rm k}(\underline{\bs}_0)_2$. Therefore, we can rewrite (\ref{l1_mmv_error_1}) as
\begin{align}
\label{l1_mmv_error_2}
\|\underline{\bs} - \underline{\bs}_0\|_1 \leq \frac{2+2\rho_m}{1-\rho_m}\sqrt{N} \sigma_{\rm k}(\underline{\bs}_0)_2 + \frac{4\tau_m}{1-\rho_m}{M}\sigma
\end{align}
where for the second part of the sight hand side of (\ref{l1_mmv_error_1}) we have used (\ref{w_bound_3}), i.e., $\|\bw(\omega)\|_2\leq M\sigma$.
Using (\ref{sigma_mmv_k_2}) we can describe (\ref{l1_mmv_error_2}) as
\begin{align}
\label{l1_mmv_error_3}
\|\underline{\bs} - \underline{\bs}_0\|_1 \leq \frac{(1+\rho_m)}{(1-\rho_m)\sqrt{\rm k}} \sqrt{N} \|\underline{\bs}_0\|_1 + \frac{4\tau_m}{1-\rho_m}M\sigma.
\end{align}
Using $\|\underline{\bs} - \underline{\bs}_0\|_{\infty}\leq \|\underline{\bs} - \underline{\bs}_0\|_1$ and $\|\underline{\bs}_0\|_1 \leq N \|\underline{\bs}_0\|_{\infty}$ we can express (\ref{l1_mmv_error_3}) as
\begin{align}
\label{l1_mmv_error_4}
\|\underline{\bs} - \underline{\bs}_0\|_{1} \leq \frac{(1+\rho_m)}{(1-\rho_m)\sqrt{\rm k}} \sqrt{N} N \|\underline{\bs}_0\|_{\infty} + \frac{4\tau_m}{1-\rho_m}M\sigma
\end{align}
Let $\boldsymbol{\delta}_m \in \mathcal{N}$, where  $\mathcal{N}$ is the null space of the matrix $\tilde{\mathbf{\Phi}}(\omega ;v)$. We further define the set $I_m$ to be a set of the indices of the nonzero elements of $\underline{\bu}_0$, i.e., $I_m={\rm supp}(\underline{\bu}_0)$, where $\underline{\bu}_0$ has been given in (\ref{model_sparse_mmv_noiseless}).  Then we partition $\boldsymbol{\delta}_m$ as $\underline{\boldsymbol{\delta}}=\underline{(\boldsymbol{\delta}}_{I_m},\underline{\boldsymbol{\delta}}_{I^c_m})$, where $I^c_m$ is the complement of $I_m$. The matrix $\tilde{\mathbf{\Phi}}(\omega ;v)$ is said to satisfy the stable  null space property with constant $0< \rho_m < 1$ if \cite{Foucart,Donoho,Donoho1}
\begin{align}
\label{null-space_property_mmv_1}
\|\underline{\boldsymbol{\delta}}_{I_m} \|_1\leq \rho_m \|\underline{\boldsymbol{\delta}}_{I^c_m}\|_1.
\end{align}
We manage to find $\rho_m$. To accomplish this goal we define $\bv_m=\tilde{\mathbf{\Phi}}_s(\omega ;v)I_m = \tilde{\mathbf{\Phi}}_{ns}(\omega ;v)I^c_m $. Then we can write the following
\begin{align}
\label{null-space_property_mmv_2}
\|\underline{\boldsymbol{\delta}}_{I_m}\|_1 \leq \sqrt{k}\|\underline{\boldsymbol{\delta}}_{I_m}\|_2.
\end{align}
We can further write
\begin{align}
\label{null-space_property_mmv_3}
\|\bv_m\|_2 = \|\tilde{\mathbf{\Phi}}_s(\omega ;v)\underline{\boldsymbol{\delta}}_{I_m}\|_2 \geq \underline{\sigma}_{{\min}_s} \|\underline{\boldsymbol{\delta}}_{I_m}\|_2
\end{align}
where $\underline{\sigma}_{{\min}_s}$ stands for the minimum singular value of the matrix $\tilde{\mathbf{\Phi}}_s(\omega ;v)$. We also know that $\underline{\sigma}_{{\min}_s} =\|\tilde{\mathbf{\Phi}}_s(\omega ;v)\|_2$, hence using (\ref{null-space_property_mmv_3}) the relation (\ref{null-space_property_mmv_2}) is expressed  as
\begin{align}
\label{null-space_property_mmv_4}
\|\underline{\boldsymbol{\delta}}_{I_m}\|_1 \leq \sqrt{k} \frac{\|\bv_m\|_2}{\|\tilde{\mathbf{\Phi}}_s(\omega ;v)\|_2} \leq \sqrt{k} \frac{\|\bv_m\|_1}{\|\tilde{\mathbf{\Phi}}_s(\omega ;v)\|_2}.
\end{align}
To connect $\|\bv_m\|_1$ to $\|\underline{\boldsymbol{\delta}}_{I^c_m}\|_1$ we use $\bv_m=\tilde{\mathbf{\Phi}}_s(\omega ;v)\underline{\boldsymbol{\delta}}_{I_m} = \tilde{\mathbf{\Phi}}_{ns}(\omega ;v)\underline{\boldsymbol{\delta}}_{I^c_m}$ which upon taking norm-1 from both sides we obtain
\begin{align}
\label{null-space_property_mmv_5}
\|\bv_m\|_1= \|\tilde{\mathbf{\Phi}}_{ns}(\omega ;v)\underline{\boldsymbol{\delta}}_{I^c_m}\|_1 \leq \|\tilde{\mathbf{\Phi}}_{ns}(\omega ;v)\|_1\|\underline{\boldsymbol{\delta}}_{I^c_m}\|_1.
\end{align}
Using (\ref{null-space_property_mmv_5}) the modified version of (\ref{null-space_property_mmv_4}) can be described as
\begin{align}
\label{null-space_property_mmv_6}
\|\underline{\boldsymbol{\delta}}_{I_m}\|_1 \leq \sqrt{k} \frac{\|\tilde{\mathbf{\Phi}}_{ns}(\omega ;v)\|_1}{\|\tilde{\mathbf{\Phi}}_s(\omega ;v)\|_2}\|\underline{\boldsymbol{\delta}}_{I^c_m}\|_1.
\end{align}
Therefore, based on (\ref{null-space_property_mmv_6}), the parameter $\rho$ in (\ref{null-space_property_mmv_1}) is given as
\begin{align}
\label{rho_mmv_1}
 \rho_m = \sqrt{k} \frac{\|\tilde{\mathbf{\Phi}}_{ns}(\omega ;v)\|_1}{\|\tilde{\mathbf{\Phi}}_s(\omega ;v)\|_2}.
\end{align}
Instead of $\underline{\bs}$ and $\underline{\bs}_0$ we can work with $\underline{\bs}^{\prime}=\displaystyle \frac{\underline{\bs}}{\max\{\underline{\bs}\}}$ and $\underline{\bs}^{\prime}_0= \displaystyle \frac{\underline{\bs}_0}{\max\{\underline{\bs}_0\}}$. As a result $\|\underline{\bs}^{\prime}_0\|_{\infty}=1$ and we can rewrite (\ref{l1_mmv_error_4}) as
\begin{align}
\label{l1_ssv_error_mmv_5}
\|\underline{\bs}^{\prime} - \underline{\bs}_0^{\prime}\|_{1} \leq \frac{(1+\rho_m)}{(1-\rho_m)\sqrt{\rm k}} \sqrt{N} N  + \frac{4\tau_m}{1-\rho_m}M\sigma.
\end{align}
Furthermore, using (\ref{rho_mmv_1}) the error bound given in (\ref{l1_ssv_error_mmv_5}) can be described as
\begin{align}
\label{l1_ssv_error_mmv_6}
\|\underline{\bs}^{\prime} - \underline{\bs}_0^{\prime}\|_{1} & \leq \frac{\left(\|\tilde{\mathbf{\Phi}}_s(\omega ;v)\|_2+\sqrt{k}\|\tilde{\mathbf{\Phi}}_{ns}(\omega ;v)\|_1\right) \sqrt{N/k} N }{\left(\|\tilde{\mathbf{\Phi}}_s(\omega ;v)\|_2-\sqrt{k}\|\tilde{\mathbf{\Phi}}_{ns}(\omega ;v)\|_1\right)}
\nonumber \\ & + \frac{4\tau_m\|\tilde{\mathbf{\Phi}}_s(\omega ;v)\|_2 M\sigma}{\left(\|\tilde{\mathbf{\Phi}}_s(\omega ;v)\|_2-\sqrt{k}\|\tilde{\mathbf{\Phi}}_{ns}(\omega ;v)\|_1\right)}.
\end{align}}\label{error_l1_MIMO}

{\subsection{Multiple Input Multiple Output for Unknown Velocity}}
In this subsection we address the problem of the MIMO-UV imaging. We start by introducing a new dictionary matrix $\mathbf{\Psi}(\omega)\in \mathbb{C}^{M^2\times \tilde{R}N}$ as
 \begin{align}
\label{psi}
\mathbf{\Psi}(\omega) \triangleq [ \tilde{\mathbf{\Phi}}(\omega ;v_1)\;\tilde{\mathbf{\Phi}}(\omega ;v_2)\;\cdots\;\tilde{\mathbf{\Phi}}(\omega ;v_{\tilde{R}})]
\end{align}
where each $\tilde{\mathbf{\Phi}}(\omega ;v_q) \in \mathbb{C}^{M^2\times N}$, for {$q \in \{ 1,2,\cdots,\tilde{R}\}$}, is given in { (\ref{phi_mmv})}, $\{v_q\}_{q =1}^{\tilde R}$  is the set of possible values for the wave velocity in the ROI, and $\tilde{R}$ is the number of the different velocities that we consider.
 Therefore, we can cast the MIMO-UV problem as a sparse signal representation problem given as{
\begin{align}
 \label{model_mmv_vel}
 {\by}(\omega) = \mathbf{\Psi}(\omega)\bar{\bs}\  + {\bw}(\omega).
 \end{align}}
In (\ref{model_mmv_vel}), the vector {$\bar{\bs} \in \mathbb{R}^{N \tilde{R} \times 1}$} is given as
\begin{align}
\bar{\bs} = [ \tilde{\bs}^T_1\;  \tilde{\bs}^T_2\;\cdots\; \tilde{\bs}^T_{\tilde{R}}]^T
\end{align}
where all $\tilde{\bs}_q$ are zero unless when $v_q = v$. When $v_q = v$ for some $q$, the $l^{\rm th}$ element of vector $\tilde{\bs}_q \in \mathbb{R}^{N\times 1}$ is nonzero if there is a scatterer located at the $(i,j)^{\rm th}$ pixel, where $i= \lfloor\frac{l-1}{n_x}\rfloor+1$ and $ j=l-(i-1)n_x $, and it is zero otherwise. In this case, each nonzero element of {$\tilde{\bs}_q$} is one of the entries of {$\underline{\boldsymbol{\rho}}\in \mathbb{R}^{L \times 1} $}, given in (\ref{ref_coeff_mmv}).  Based on (\ref{model_mmv_vel}), the  corresponding $\ell_1$-norm  minimization problem is described as {
\begin{align}
 \label{l1_mmv_vel}
 \rm{\min_{\bar{\bs}}}. \ & \|\bar{\bs}\|_1 \  \nonumber \\
 \rm{subject \ to} \ & \|\by(\omega) - \mathbf{\Psi}(\omega) \bar{\bs}\|_2\le \beta_v \nonumber \\
 & \bar{\bs} \succeq 0, \;\;\; \bar{\bs}\in \mathbb{R}_{+}^{\tilde{R}N \times 1}
 \end{align}}
{where $\beta_v$ is the regularization parameter and is chosen such that the probability of the norm of the noise vector being larger than $\beta_v$ is small \cite{Malioutov_journal,Thesis}. The optimization problem given in (\ref{l1_mmv_vel}) can easily be solved using CVX software package \cite{cvx}. In fact CVX casts (\ref{l1_mmv_vel}) as a SOCP optimization problem that can be handled efficiently even for large scale problems. }\label{beta_v}
In reality due to modeling errors including grid resolution and noise $\tilde{\bs}_{{q}}$ may not be zero when $v_{q} \ne v$. Nevertheless, it is very likely that the entries of {$\tilde{\bs}_q$} are much smaller compared to the case when $v_q = v$. Hence we can find the value of $q$ for which  {$\tilde{\bs}_q$} has the largest $\ell_2$ norm. That is if we define $\hat{q} \triangleq \displaystyle {\arg \max_{1 \leq q \leq \tilde{R}}} \|\tilde{\bs}_q\|_2$, we introduce $\tilde \bs_{\hat q}$ as the vectorized version of the image of the ROI. Indeed, the $l^{\rm th}$ element of vector $\tilde{\bs}_{\hat{q}} \in \mathbb{C}^{N\times 1}$ is nonzero if there is a scatterer located at the $(i,j)^{\rm th}$ pixel, where $i= \lfloor\frac{l-1}{n_x}\rfloor+1$ and $ j=l-(i-1)n_x $, and it is zero otherwise. Each nonzero element of $\tilde{\bs}_{\hat{q}}$   is one of the entries of {$\underline{\boldsymbol{\rho}}\in \mathbb{C}^{L \times 1}$}, given in (\ref{ref_coeff_mmv}). Finally by reshaping $\tilde{\bs}_{\hat{q}}$ into an $n_z \times n_x$ matrix, we obtain  the final image.

{
\paragraph{$\ell_1$-norm Minimization Solver for the MIMO-UV Case}
To solve (\ref{l1_mmv_vel}), SeDuMi  solver uses the barrier method or path following method \cite{Boyd_book}.  The inequalities in (\ref{l1_mmv_vel}) define the positive orthant in $\mathbb{R}^{N \tilde{R} \times 1}$. This can be equipped with the following self-concordant barrier (in fact this is $\nu$-self-concordant barrier)
\begin{align}
\label{barrier_mmv_uv}
F_v(\bar{\bs}) = -\sum\limits_{i = 1}^{N\tilde{R}} \log(\bar{\bs}(i)),  \;\;\; \nu_v=N\tilde{R}.
\end{align}
This barrier is called the standard logarithmic barrier for $\mathbb{R}_{+}^{N\tilde{R}\times 1}$. In fact the standard logarithmic barrier is optimal for $\mathbb{R}_{+}^{N\tilde{R}\times 1}$ \cite[Theorem 4.3.2]{Nesterov}.
The complexity estimate of solving (\ref{l1_mmv_vel}) using SeDuMi solver is $O(\sqrt{N\tilde{R}}\ln(\frac{1}{\epsilon_v}))$ \cite{sturm}. The parameter $\epsilon_v$ is given as $|\bar{\bs} - \bar{\bs}^{*}| \leq \epsilon_v$, where $\bar{\bs}^{*}$ stands for the optimal value of $\bar{\bs}$, in (\ref{l1_mmv_vel}).}\label{cvx_MIMO_UV}

\section{Experimental Examples}\label{sec:exp}
\begin{figure}
\centerline{
\includegraphics[height=3cm,width=5.5cm]{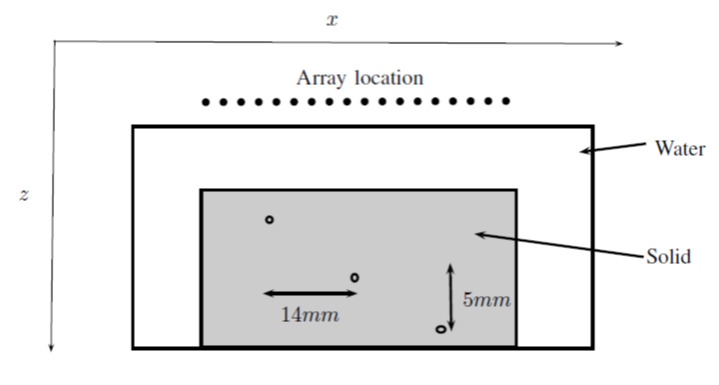}
\hspace{0.1cm}
}
\vspace*{0.1cm}
\caption{Test setup.
\label{fig:schematic}}
\end{figure}

In this section, we apply the DAS beamformer, the MUSIC technique, the Capon method as well as different sparsity based techniques that we have discussed in the paper to experimental data gathered from a test sample immersed in water, and generate the corresponding images and discuss their differences. The setup has been shown in Fig.~\ref{fig:schematic}, where a solid object with three holes is immersed in water. The horizontal and vertical distances between the holes are 14 mm and 5 mm, respectively. The specifications for the array have been summarized in Table \ref{table1}.

\begin{table}
\caption{Array parameters specification. \label{table1}}
\begin{center}
\begin{tabular}{|l|l|}\hline
Array parameters & Value \\\hline
Element type & rectangular \\\hline
Number of elements & 64  \\\hline
Element Pitch & 0.6 mm \\\hline
Element width & 0.53 mm \\\hline
Element length & 0.012 m \\\hline
Center Frequency & 5 MHz\\\hline
Sampling frequency & 100 MHz \\\hline
Bandwith &5MHz\\\hline
Speed of wave in water &1482 m/s\\\hline
Speed of wave in solid & 6400 m/s \\\hline

\end{tabular}
\end{center}
\end{table}

\begin{figure*}

\psfrag{x [mm]}[ct][cb]{$x$ [mm]}

\psfrag{z [mm]}[rb][ct]{$z$ [mm]}
\psfrag{y [mm]}[rb][ct]{$z$ [mm]}

\psfrag{Depth}[c][c]{\tiny{\rm{Depth}}}

\psfrag{Lateral Position}[ct]{\tiny{\rm{Lateral Position}}}

\psfrag{DAS}[c][c]{\tiny Normalized $\mathcal{I}_{\rm{DAS}}$  }

\psfrag{Capon}[c][c]{\tiny Normalized $\mathcal{I}_{\rm{Capon}}$ }

\psfrag{MUSIC}[c][c]{\tiny Normalized  $\mathcal{I}_{\rm{MUSIC}}$ }
\psfrag{l1-SMV}[c][c]{\tiny Normalized  $\mathcal{I}_{\rm{l1-SIMO}}$  }

\psfrag{l1-MMV}[c][c]{\tiny Normalized $\mathcal{I}_{\rm{l1-MIMO}}$ }

\psfrag{l1-MMV-UV}[c][c]{\tiny Normalized $\mathcal{I}_{\rm{l1-MIMO-UV}}$ }

\psfrag{Capon Image t}{}
\psfrag{MUSIC Image t}{}
\psfrag{DAS Image t}{}
\psfrag{L1-SIMO Image t}{}
\psfrag{L1-MIMO Image t}{}
\psfrag{L1-MIMO-UV Image t}{}

\centerline{
\includegraphics[height=4.5cm,width=5.5cm]{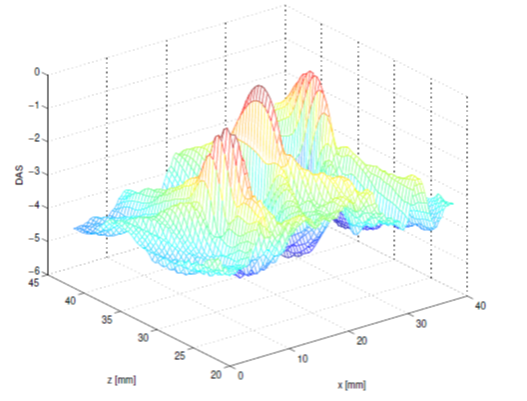}
\hspace{0.5cm}
\includegraphics[height=4.5cm,width=5.5cm]{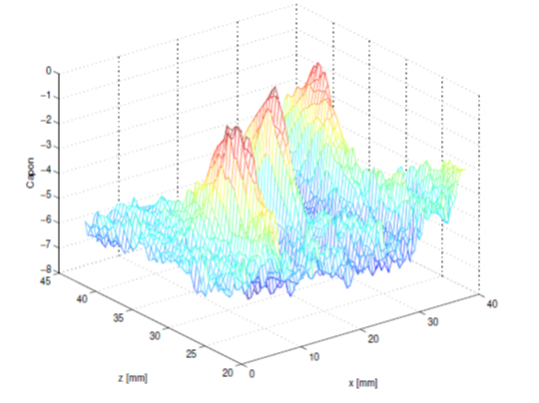}
\hspace{0.5cm}
\includegraphics[height=4.5cm,width=5.5cm]{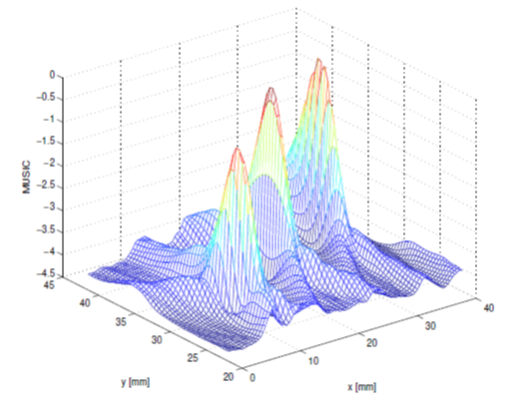}
}

\centerline{(a)\hspace*{5.5cm}(b)\hspace*{5.5cm}(c)}
\vspace*{0.2in}

\centerline{
\includegraphics[height=4.5cm,width=5.5cm]{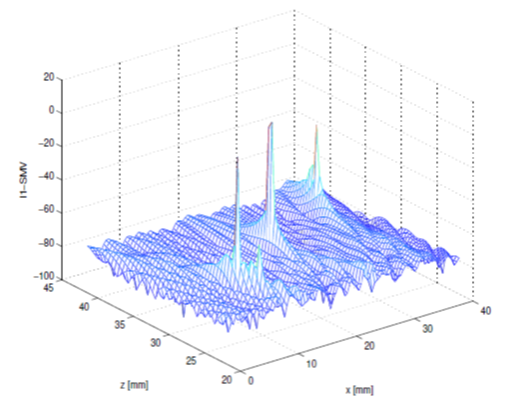}
\hspace{0.5cm}
\includegraphics[height=4.5cm,width=5.5cm]{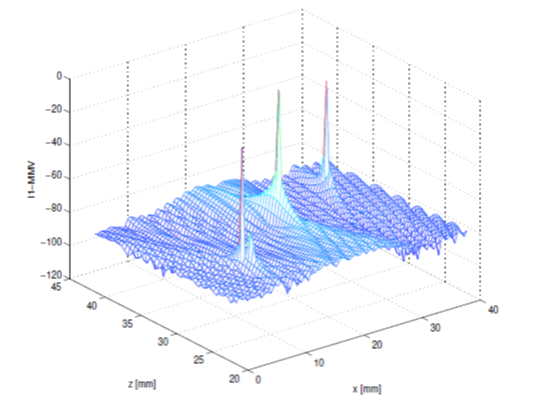}
\hspace{0.5cm}
\includegraphics[height=4.5cm,width=5.5cm]{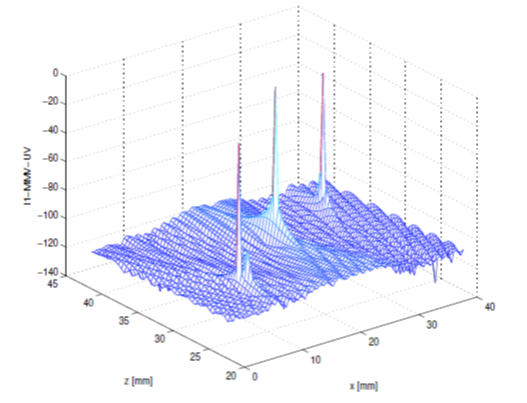}
}
\centerline{(d)\hspace*{5.5cm}(e)\hspace*{5.5cm}(f)}

\caption{The 3D image representation for (a) the DAS beamformer technique based image using (\ref{DAS}), (b) the Capon technique based image using (\ref{Capon_all_freq}), (c) the Music method based image using (\ref{music_all_freq}),  (d) the $\ell_1$-norm minimization based image for the SIMO case  using (\ref{l1_SMV}), (e) the $\ell_1$-norm minimization based image for the MIMO case  using (\ref{l1_mmv}), (f) the $\ell_1$-norm minimization based image for the MIMO-UV case using (\ref{l1_mmv_vel}).
\label{fig:2D}}

\end{figure*}

The number of the snapshots is equal to the number of the transducers which is 64. The sample covariance matrix in (\ref{covariance_matrix}) is calculated based on all the snapshots. The size of each bin of the FFT that has been used to map our data into frequency domain is $41.67$ KHz. A total number of 120 bins are used for the DAS beamformer, the MUSIC technique, and the Capon method within the whole $5$ MHz bandwidth of our probing signal. Fig.~\ref{fig:2D} show the 3D  images. As can be seen from Fig.~\ref{fig:2D}-(a), the DAS technique based on (\ref{DAS}), suffers from low resolution and high sidelobe levels. Figs.~\ref{fig:2D}-(b) and (c) illustrate the results of {the MUSIC method and the Capon technique} based on (\ref{music_all_freq}) and (\ref{Capon_all_freq}), respectively. The size of the signal subspace for {the} MUSIC technique, i.e., $L$, has been set to $3$. The loading parameter $\kappa$ for {the} Capon technique has been set to $10$ to $12$ dB above the estimated noise {level}. The estimate of the noise power is obtained as {$\frac{\sum_{r=L+1}^{M}\sigma_{r}}{M-L-1}$}\label{kappa} \cite{diagonal_loading_1,diagonal_loading_2,diagonal_loading_3}, where $\sigma_{r}$ is the $r^{\rm th}$ largest eigenvalue of the sample covariance matrix and $L$ has been set to 3. {The poor results for both techniques are partly due to relatively low number of snapshots as only 64 snapshots are not sufficient to obtain a good estimate of the covariance matrix. As to the Capon method, the other issue is the parameter $\kappa$. Since the Capon technique is sensitive to $\kappa$ and since there is only a rule of thumb for choosing $\kappa$, which says $\kappa$ should be 10 to 12 dB above the noise level, therefore the performance of the Capon method changes by $\kappa$.}\label{Capon_DAS}

Compared to sparse signal representation based techniques their resolution are much lower. Fig.~\ref{fig:2D}-(d) shows the result for the SIMO case based on (\ref{l1_SMV}). Unlike {the DAS beamformer,  the MUSIC technique and the Capon method}, for sparse signal representation based techniques, we have used one frequency bin corresponding to the center frequency which is $5$ MHz. {Furthermore, for the SIMO case only one measurement vector, corresponding to the case that one of the transducers has fired, has been used.} The image shown in Fig.~\ref{fig:2D}-(d) corresponds to the case when the $30^{\rm th}$ transducer has fired the signal. Using only one frequency bin and one snapshot, Fig.~\ref{fig:2D}-(d) shows the superiority of the SIMO based method over {the DAS beamformer, the MUSIC technique and the Capon method}. As can be seen from Fig.~\ref{fig:2D}-(d), the sidelobe levels are more than $80$ dB lower than the peaks. Fig.~\ref{fig:2D}-(e) illustrates the result for the MIMO case based on (\ref{l1_mmv}). Similar to the SIMO case, only one frequency bin corresponding to the center frequency has been used for the the MIMO based image. From Fig.~\ref{fig:2D}-(e), we see that sidelobe levels are more than $110$ dB lower than the peaks.

For all the results so far we have assumed that the propagation velocity of the ultrasonic wave in the specimen is known. Fig.~\ref{fig:2D}-(f) shows the result for the MIMO-UV based image based on (\ref{l1_mmv_vel}). The set of proposed velocities we have chosen is $\{5800, \; 6000\;, 6400,\; 6600\}$ m/s. Our motivation for that comes from the fact that the propagation velocity of ultrasonic waves in this frequencies fall in this range. Of course there is no limitation here and the ranges of the proposed velocities can cover any values. Fig.~\ref{fig:2D}-(f) also illustrates the same superiority that we mentioned for Fig.~\ref{fig:2D}-(d)-(e).

\section{Conclusions}\label{conclusion}

\title{Super Resolution Ultrasonic Imaging of Two-Layer Objects using Sparsity}

 \maketitle

{In this paper, we modeled wave refraction which occurs at the interface of two media using Huygens principle and used this model to develop sparse signal representation based imaging techniques for a two-layer object immersed in water.
Relying on the fact that the image of interest is sparse,  we cast such an array based imaging problem as a sparse signal recovery problem and developed two types of imaging methods, one method uses only one transducer to illuminate the region of interest and the second method relies on all transducers to transmit ultrasonic waves into the material under test. We discussed greedy based algorithms as well as the $\ell_1$-norm minimization based method as well known candidates to recover the sparse signal. We then mentioned the reasons why the $\ell_1$-norm minimization based method is more popular than  greedy based algorithms.  We further obtained an upper bound for the error for both  greedy based algorithms as well as the $\ell_1$-norm minimization based method  in both the SIMO case and the MIMO case. We further extended our work to a scenario where the propagation velocity of the wave in the object under test is not known precisely. We discussed the software packages and the techniques that they use to solve the $\ell_1$-norm minimization based method for the SIMO, the MIMO and the MIMO-UV cases. We then provided the computational complexity for each of the SIMO, the MIMO and the MIMO-UV cases. We applied  our methods to experimental data gathered from a solid test sample immersed in water and showed that our sparse based techniques outperform the conventional methods available in the literature.
The conclusion is that sparse signal representation based techniques outperform the DAS beamformer, the MUSIC technique and the Capon method by a large margin in higher resolution, lower sidelobe levels and being less sensitive to SNR and correlated targets.
}

\bibliographystyle{IEEEtran}
\bibliography{Biblio}

@ARTICLE{RMS_sciesmology,
author={Schneider, W.A.},
journal={Proceedings of the IEEE},
title={The common depth point stack},
year={1984},
volume={72},
number={10},
pages={1238-1254},
doi={10.1109/PROC.1984.13014},
ISSN={0018-9219},}

@ARTICLE{DAS,
 author={Wilcox, P.D.},
 journal={ IEEE Transactions on, Ultrasonics, Ferroelectrics, and Frequency Control},
 title={Omni-directional guided wave transducer arrays for the rapid inspection of large areas of plate structures},
 year={2003},
 month={June},
 volume={50},
 number={6},
 pages={699-709},
 doi={10.1109/TUFFC.2003.1209557},
 ISSN={0885-3010},}

@ARTICLE{ML_DAS,
author={M. H. Skjelvareid and Y. Birkelund},
journal={Proceedings of the ASME},
title={Ultrasound imaging using multilayer synthetic aperture focusing},
year={2010},
volume={5},
number={PVP2010-25338},
pages={379-387},
doi={10.1115/PVP2010-25338},
ISBN={978-0-7918-4924-8},}

@Book{Rayleigh,
author={D. H. Johnson and D. E. Dudgeon},
title={ Array Signal Processing - Concept and Techniques},
year={1993},
publisher= {Prentice Hall},
address={Prentice Hall },}

@ARTICLE{Music,
author={Schmidt, R.O.},
journal={ IEEE Transactions on  Antennas and Propagation},
title={Multiple emitter location and signal parameter estimation},
year={1986},
volume={34},
number={3},
pages={276-280},
doi={10.1109/TAP.1986.1143830},
ISSN={0018-926X},}

@ARTICLE{Music_stoica,
author={Stoica, Petre and Nehorai Arye},
journal={ IEEE Transactions on  Acoustics, Speech and Signal Processing},
title={MUSIC, maximum likelihood, and Cramer-Rao bound},
year={1989},
volume={37},
number={5},
pages={720-741},
doi={10.1109/29.17564},
ISSN={0096-3518},}

@Book{Stoica_book,
author = {P. Stoica and R. L. Moses},
title = {Spectral Analysis of Signals},
publisher = {Prentice-Hal},
year = {2005},
address = {Upper Saddle River, NJ},}

@ARTICLE{capon_o,
author={Capon, J.},
journal={Proceedings of the IEEE},
title={High-resolution frequency-wavenumber spectrum analysis},
year={1969},
volume={57},
number={8},
pages={1408-1418},
doi={10.1109/PROC.1969.7278},
ISSN={0018-9219},}

@PHDTHESIS{Thesis,
    AUTHOR = {D. M. Malioutov},
    TITLE = {A Sparse Signal Reconstruciton Perspective for Source Localization With Sensor Arrays},
    SCHOOL = {Mass. Inst. Technol.},
    ADDRESS = { Cambridge, MA},
    YEAR ={2003},}

@ARTICLE{Malioutov_journal,
author={Malioutov, D. and Cetin, M. and Willsky, A.S.},
journal={ IEEE Transactions on  Signal Processing},
title={A sparse signal reconstruction perspective for source localization with sensor arrays},
year={2005},
volume={53},
number={8},
pages={3010-3022},
doi={10.1109/TSP.2005.850882},
ISSN={1053-587X},}

@ARTICLE{doa_sparse,
 author={Xu Xu and Xiaohan Wei and Zhongfu Ye},
 journal={ IEEE  Signal Processing Letters},
 title={DOA Estimation Based on Sparse Signal Recovery Utilizing Weighted $l_1$ -Norm Penalty},
 year={2012},
 month={March},
 volume={19},
 number={3},
 pages={155-158},
 doi={10.1109/LSP.2012.2183592},
 ISSN={1070-9908},}

@ARTICLE{SAR_application,
author={Cetin, M. and Karl, W.C.},
journal={ IEEE Transactions on   Image Processing},
title={Feature-enhanced synthetic aperture radar image formation based on nonquadratic regularization},
year={2001},
volume={10},
number={4},
pages={623-631},
doi={10.1109/83.913596},
ISSN={1057-7149},}

@ARTICLE{image_application,
author={Charbonnier, P. and Blanc-Feraud, L. and Aubert, G. and Barlaud, M.},
journal={ IEEE Transactions on  Image Processing},
title={Deterministic edge-preserving regularization in computed imaging},
year={1997},
volume={6},
number={2},
pages={298-311},
doi={10.1109/83.551699},}

@article{spectrum_application,
author={Jeffs, B.D.},
journal={ IEEE International Conference on Acoustics, Speech and Signal Processing},
title={Sparse inverse solution methods for signal and image processing applications},
year={1998},
volume={3},
pages={1885-1888 vol.3},
doi={10.1109/ICASSP.1998.681832},
ISSN={1520-6149},}

@article{application_array_pro,
author={Fuchs, J-J},
journal ={ IEEE International Conference on  Acoustics, Speech, and Signal Processing,  ICASSP-96.},
title={Linear programming in spectral estimation. Application to array processing},
year={1996},
volume={6},
pages={3161-3164 vol. 6},
doi={10.1109/ICASSP.1996.550547},
ISSN={1520-6149},}

@ARTICLE{basis_pusuit,
author={S. S. Chen and D. L. Donoho and M. A. Saunders},
journal={SIAM J. Scientific Computing},
title={Atomic decomposition by basis pursuit},
year={1998},
volume={20},
number={1},
pages={33-61},}

@ARTICLE{lasso,
author={W. J.  Fu},
journal={Journal of Computational and Graphical Statistics},
title={Penalized regressions the bridge versus the lasso},
year={1998},
volume={7},
number={3},
 pages={397-416},}

@ARTICLE{Zijan_Kruskal,
 author={Zijian Tang and Blacquiere, G. and Leus, G.},
 journal={ IEEE Transactions on  Signal Processing},
 title={Aliasing-Free Wideband Beamforming Using Sparse Signal Representation},
 year={2011},
 month={July},
 volume={59},
 number={7},
 pages={3464-3469},
 doi={10.1109/TSP.2011.2140108},
 ISSN={1053-587X},}

@ARTICLE{Chen_Kruskal,
 author={Jie Chen and Huo, X.},
 journal={ IEEE Transactions on  Signal Processing},
 title={Theoretical Results on Sparse Representations of Multiple-Measurement Vectors},
 year={2006},
 month={Dec},
 volume={54},
 number={12},
 pages={4634-4643},
 doi={10.1109/TSP.2006.881263},
 ISSN={1053-587X},}

@article{DoBa,
 author = {Do Ba, Khanh and Indyk, Piotr and Price, Eric and Woodruff, David P.},
 title = {Lower Bounds for Sparse Recovery},
 journal = {Proceedings of the Twenty-first Annual ACM-SIAM Symposium on Discrete Algorithms},
 series = {SODA '10},
 year = {2010},
 isbn = {978-0-898716-98-6},
 location = {Austin, Texas},
 pages = {1190--1197},
 numpages = {8},
  acmid = {1873696},
 publisher = {Society for Industrial and Applied Mathematics},
 address = {Philadelphia, PA, USA},
}

@ARTICLE{MMV_Hyder,
 author={Hyder, M.M. and Mahata, K.},
 journal={IEEE Transactions on  Signal Processing},
 title={Direction-of-Arrival Estimation Using a Mixed $\ell _{2,0}$  Norm Approximation},
 year={2010},
 month={Sept},
 volume={58},
 number={9},
 pages={4646-4655},
 doi={10.1109/TSP.2010.2050477},
 ISSN={1053-587X},}

@ARTICLE{MMV_Cotter,
 author={Cotter, S.F. and Rao, B.D. and Engan, K. and Kreutz-Delgado, K.},
 journal={ IEEE Transactions on  Signal Processing},
 title={Sparse solutions to linear inverse problems with multiple measurement vectors},
 year={2005},
 month={July},
 volume={53},
 number={7},
 pages={2477-2488},
 doi={10.1109/TSP.2005.849172},
 ISSN={1053-587X},}

@ARTICLE{MMV_Eldar,
 author={Eldar, Y.C. and Mishali, M.},
 journal={ IEEE Transactions on   Information Theory},
 title={Robust Recovery of Signals From a Structured Union of Subspaces},
 year={2009},
 month={Nov},
 volume={55},
 number={11},
 pages={5302-5316},
 doi={10.1109/TIT.2009.2030471},
 ISSN={0018-9448},}

@ARTICLE{Eldar-Ultrasound,
 author={Wagner, N. and Eldar, Y.C. and Friedman, Z.},
 journal={ IEEE Transactions on  Signal Processing},
 title={Compressed Beamforming in Ultrasound Imaging},
 year={2012},
 month={Sept},
 volume={60},
 number={9},
 pages={4643-4657},
 doi={10.1109/TSP.2012.2200891},
 ISSN={1053-587X},}

@Article{sar,
AUTHOR = {Guarneri, Giovanni Alfredo and Pipa, Daniel Rodrigues and Junior, Flávio Neves and de Arruda, Lúcia Valéria Ramos and Zibetti, Marcelo Victor Wüst},
TITLE = {A Sparse Reconstruction Algorithm for Ultrasonic Images in Nondestructive Testing},
JOURNAL = {Sensors},
VOLUME = {15},
YEAR = {2015},
NUMBER = {4},
PAGES = {9324},
PubMedID = {25905700},
ISSN = {1424-8220},
DOI = {10.3390/s150409324},
}

@ARTICLE{Shahrokh, 
 author={Hamidi, S. and Shahbazpanahi, S.}, 
 journal={ IEEE Transactions on  Signal Processing}, 
 title={Sparse Signal Recovery based Imaging in the Presence of Mode Conversion with Application to Non-Destructive Testing}, 
 year={2015}, 
 volume={PP}, 
 number={99}, 
 doi={10.1109/TSP.2015.2486742}, 
 ISSN={1053-587X}, 
 month={Oct},}

@Book{Huygens,
author = { Joseph W. Goodman},
title = {Introduction to Fourier Optics},
publisher = {McGraw-Hill},
year = {1996},}

@ARTICLE{Donoho,
author={Donoho, D.L.},
journal={IEEE Transactions on  Information Theory},
title={Compressed sensing},
year={2006},
volume={52},
month={April},
pages={1289-1306},
doi={10.1109/TIT.2006.871582},
ISSN={0018-9448},
}

@Book{Boyd_book,
author = {Stephen Boyd and Lieven Vandenberghe},
title = {Convex Optimization},
publisher = {Cambridge University Press},
year = {2004},
address = {Cambridge},}

@ARTICLE{diagonal_loading_1,
 author={ShahbazPanahi, S. and Gershman, A.B. and Zhi-Quan Luo and Kon Max Wong},
 journal={ IEEE Transactions on  Signal Processing},
 title={Robust adaptive beamforming for general-rank signal models},
 year={2003},
 month={Sept},
 volume={51},
 number={9},
 pages={2257-2269},
 doi={10.1109/TSP.2003.815395},
 ISSN={1053-587X},}

@ARTICLE{diagonal_loading_2,
 author={Carlson, B.D.},
 journal={IEEE Transactions on  Aerospace and Electronic Systems},
 title={Covariance matrix estimation errors and diagonal loading in adaptive arrays},
 year={1988},
 month={Jul},
 volume={24},
 number={4},
 pages={397-401},
 doi={10.1109/7.7181},
 ISSN={0018-9251},}

@ARTICLE{diagonal_loading_3,
 author={Bell, K.L. and Ephraim, Y. and Van Trees, H.L.},
 journal={IEEE Transactions on  Signal Processing},
 title={A Bayesian approach to robust adaptive beamforming},
 year={2000},
 month={Feb},
 volume={48},
 number={2},
 pages={386-398},
 doi={10.1109/78.823966},
 ISSN={1053-587X},}

@ARTICLE{NasimPaper2,
author={Moallemi, N. and ShahbazPanahi, S.},
journal={IEEE Transactions on Signal Processing},
title={A New Model for Array Spatial Signature for Two-Layer Imaging With Applications to Nondestructive Testing Using Ultrasonic Arrays},
year={2015},
month={May},
volume={63},
pages={2464-2475},
doi={10.1109/TSP.2015.2403273},
ISSN={1053-587X},}

@ARTICLE{Elad,
    author = {David L. Donoho and Michael Elad and Vladimir N. Temlyakov},
    title = {Stable recovery of sparse overcomplete representations in the presence of noise},
    journal = {IEEE TRANS. INFORM. THEORY},
    year = {2006},
    volume = {52},
    number = {1},
    pages = {6-18},
}

@misc{cvx,
  author       = {Michael Grant and Stephen Boyd},
  title        = {{CVX}: Matlab Software for Disciplined Convex Programming, version 2.1},
  month        = {mar},
  year         = {2014},
}

@book{Matrix,
 author = {Horn, Roger. and Johnson, Charles.},
 title= {Matrix Analysis},
 publisher = {Cambridge University Press },
 year = 1985}

@BOOK{Measure_concentration,
title = {The Concentration of Measure Phenomenon}, 
author = {Michel Ledoux },
publisher = { American Mathematical Soc. Issue 89 of Mathematical Surveys and Monographs},
year = {2005},}

@BOOK{Ledoux,
 author = {Ledoux, Michel.  and  Talagrand, Michel. },
 title= {Probability in Banach Spaces: Isoperimetry and Processes},
 publisher = {Springer},
 year = 2013}

@BOOK{Milman,
 author = { Milman, Vitali D.   and   Schechtman, Gideon.},
 title= {Asymptotic Theory of Finite Dimensional Normed Spaces: Isoperimetric Inequalities in Riemannian Manifolds (Lecture Notes in Mathematics) },
 publisher = {Springer},
 year = 2002}

@BOOK{Rudin,
 author = { Rudin, Walter.},
 title= {Principles of Mathematical Analysis  },
 publisher = {McGraw-Hill },
 year = 1976}

@BOOK{Gilles,
title = {The Volume of Convex Bodies and Banach Space Geometry},
author = {Gilles Pisier}, 
publisher = { Cambridge University Press, 1999 },
year = {1999},}

@BOOK{Mitzenmacher,
title = {Probability and Computing},
author = { M. Mitzenmacher and E. Upfal},
publisher = { Cambridge University Press},
year = {2005},}

@BOOK{GOA,
title = {Fitting Equations to Data: Computer Analysis of Multifactor Data},
author = { Daniel, Cuthbert and  Wood, Fred S. },
publisher = {John Wiley},
year = {1999},}

@article{MP,
 author = {Mallat, S.G. and Zhang, Zhifeng},
 title = {Matching Pursuits with Time-frequency Dictionaries},
 journal = {Trans. Sig. Proc.},
 issue_date = {December 1993},
 volume = {41},
 number = {12},
 month = dec,
 year = {1993},
 issn = {1053-587X},
 pages = {3397--3415},
 numpages = {19},
 doi = {10.1109/78.258082},
 acmid = {2203996},
 publisher = {IEEE Press},
 address = {Piscataway, NJ, USA},
}

@Article{OMP,
    author = {Y. C. Pati and R. Rezaiifar and Y. C. Pati R. Rezaiifar and P. S. Krishnaprasad},
    title = {Orthogonal Matching Pursuit: Recursive Function Approximation with Applications to Wavelet Decomposition},
    journal = {Proceedings of the 27 th Annual Asilomar Conference on Signals, Systems, and Computers},
    year = {1993},
    pages = {40-44},
}

@article{OGA,
year={1996},
issn={1019-7168},
journal={Advances in Computational Mathematics},
volume={5},
number={1},
doi={10.1007/BF02124742},
title={Some remarks on greedy algorithms},
publisher={Baltzer Science Publishers, Baarn/Kluwer Academic Publishers},
author={DeVore, R.A. and Temlyakov, V.N.},
pages={173-187},
}

@article{Lobo,
title = {Applications of second-order cone programming },
journal = {Linear Algebra and its Applications },
volume = {284},
number = {1–3},
pages = {193 - 228},
year = {1998},
note = {International Linear Algebra Society (ILAS) Symposium on Fast Algorithms for Control, Signals and Image Processing },
issn = {0024-3795},
author = {Miguel Sousa Lobo and Lieven Vandenberghe and Stephen Boyd and Hervé Lebret},
}

@book{Nesterov,
   title = {Introductory lectures on convex optimization : a basic course},
   author = { Nesterov, Yurii},
   series = {Applied optimization},
   publisher = {Kluwer Academic Publ.},
   address = {Boston, Dordrecht, London},
   isbn = {1-4020-7553-7},
   year = {2004},
}

@article{cvx_guide,
  author    = { Grant, M. and Boyd, S.},
  title     = {Graph implementations for nonsmooth convex programs},
  journal = {Recent Advances in Learning and Control},
  publisher = {Springer-Verlag Limited},
  pages     = {95--110},
  year      = {2008},
}

@article{sturm,
  title={USING SEDUMI 1.02, A MATLAB* TOOLBOX FOR OPTIMIZATION OVER SYMMETRIC CONESt},
  author={STURM, JOSF},
  year={2001},
}

@book{Foucart,
 author = {Foucart, Simon and Rauhut, Holger},
 title = {A Mathematical Introduction to Compressive Sensing},
 year = {2013},
 isbn = {0817649476, 9780817649470},
 publisher = { Birkhauser},
}

@article{Donoho1,
    author = {Donoho, David },
    title = {For most large underdetermined systems of equations, the minimal l1-norm near-solution approximates the sparsest near-solution},
    journal = {Comm. Pure Appl. Math},
    year = {2004},
}
\begin{IEEEbiography}[{\includegraphics[width=3cm,height=4cm,clip,keepaspectratio]{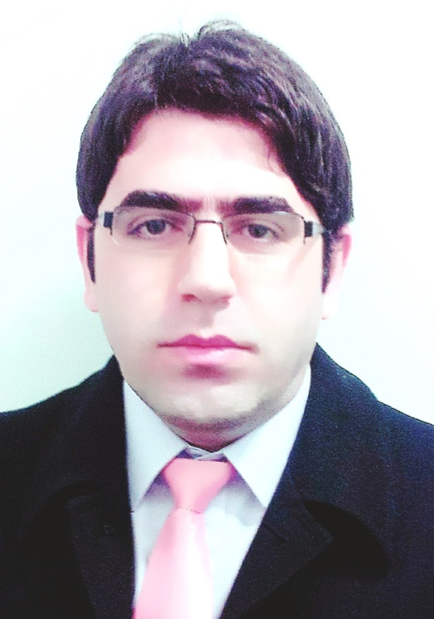}}]%
{Shahrokh Hamidi}
was born in Iran, in 1983. He received his B.Sc., M.Sc., and Ph.D. degrees all in Electrical and Computer Engineering. He is with the Faculty of Electrical and Computer
Engineering, University of Waterloo, Waterloo, ON, Canada. His research areas include statistical signal processing, image processing, wireless communication, machine learning, optimization and array processing.
\end{IEEEbiography}

\clearpage

\end{document}